\makeatletter

\font\tenmsx=msxm10
\font\sevenmsx=msxm7
\font\fivemsx=msxm5
\font\tenmsy=msym10
\font\sevenmsy=msym7
\font\fivemsy=msym5
\newfam\msxfam
\newfam\msyfam
\textfont\msxfam=\tenmsx  \scriptfont\msxfam=\sevenmsx
  \scriptscriptfont\msxfam=\fivemsx
\textfont\msyfam=\tenmsy  \scriptfont\msyfam=\sevenmsy
  \scriptscriptfont\msyfam=\fivemsy

\def\hexnumber@#1{\ifnum#1<10 \number#1\else
 \ifnum#1=10 A\else\ifnum#1=11 B\else\ifnum#1=12 C\else
 \ifnum#1=13 D\else\ifnum#1=14 E\else\ifnum#1=15 F\fi\fi\fi\fi\fi\fi\fi}

\def\msx@{\hexnumber@\msxfam}
\def\msy@{\hexnumber@\msyfam}
\mathchardef\boxdot="2\msx@00
\mathchardef\boxplus="2\msx@01
\mathchardef\boxtimes="2\msx@02
\mathchardef\square="0\msx@03
\mathchardef\blacksquare="0\msx@04
\mathchardef\centerdot="2\msx@05
\mathchardef\lozenge="0\msx@06
\mathchardef\blacklozenge="0\msx@07
\mathchardef\circlearrowright="3\msx@08
\mathchardef\circlearrowleft="3\msx@09
\mathchardef\rightleftharpoons="3\msx@0A
\mathchardef\leftrightharpoons="3\msx@0B
\mathchardef\boxminus="2\msx@0C
\mathchardef\Vdash="3\msx@0D
\mathchardef\Vvdash="3\msx@0E
\mathchardef\vDash="3\msx@0F
\mathchardef\twoheadrightarrow="3\msx@10
\mathchardef\twoheadleftarrow="3\msx@11
\mathchardef\leftleftarrows="3\msx@12
\mathchardef\rightrightarrows="3\msx@13
\mathchardef\upuparrows="3\msx@14
\mathchardef\downdownarrows="3\msx@15
\mathchardef\upharpoonright="3\msx@16

\mathchardef\downharpoonright="3\msx@17
\mathchardef\upharpoonleft="3\msx@18
\mathchardef\downharpoonleft="3\msx@19
\mathchardef\rightarrowtail="3\msx@1A
\mathchardef\leftarrowtail="3\msx@1B
\mathchardef\leftrightarrows="3\msx@1C
\mathchardef\rightleftarrows="3\msx@1D
\mathchardef\Lsh="3\msx@1E
\mathchardef\Rsh="3\msx@1F
\mathchardef\rightsquigarrow="3\msx@20
\mathchardef\leftrightsquigarrow="3\msx@21
\mathchardef\looparrowleft="3\msx@22
\mathchardef\looparrowright="3\msx@23
\mathchardef\circeq="3\msx@24
\mathchardef\succsim="3\msx@25
\mathchardef\gtrsim="3\msx@26
\mathchardef\gtrapprox="3\msx@27
\mathchardef\multimap="3\msx@28
\mathchardef\therefore="3\msx@29
\mathchardef\because="3\msx@2A
\mathchardef\doteqdot="3\msx@2B

\mathchardef\triangleq="3\msx@2C
\mathchardef\precsim="3\msx@2D
\mathchardef\lesssim="3\msx@2E
\mathchardef\lessapprox="3\msx@2F
\mathchardef\eqslantless="3\msx@30
\mathchardef\eqslantgtr="3\msx@31
\mathchardef\curlyeqprec="3\msx@32
\mathchardef\curlyeqsucc="3\msx@33
\mathchardef\preccurlyeq="3\msx@34
\mathchardef\leqq="3\msx@35
\mathchardef\leqslant="3\msx@36
\mathchardef\lessgtr="3\msx@37
\mathchardef\backprime="0\msx@38
\mathchardef\risingdotseq="3\msx@3A
\mathchardef\fallingdotseq="3\msx@3B
\mathchardef\succcurlyeq="3\msx@3C
\mathchardef\geqq="3\msx@3D
\mathchardef\geqslant="3\msx@3E
\mathchardef\gtrless="3\msx@3F
\mathchardef\sqsubset="3\msx@40
\mathchardef\sqsupset="3\msx@41
\mathchardef\trianglerighteq="3\msx@44
\mathchardef\trianglelefteq="3\msx@45
\mathchardef\bigstar="0\msx@46
\mathchardef\between="3\msx@47
\mathchardef\blacktriangledown="0\msx@48
\mathchardef\blacktriangleright="3\msx@49
\mathchardef\blacktriangleleft="3\msx@4A
\mathchardef\blacktriangle="0\msx@4E
\mathchardef\triangledown="0\msx@4F
\mathchardef\eqcirc="3\msx@50
\mathchardef\lesseqgtr="3\msx@51
\mathchardef\gtreqless="3\msx@52
\mathchardef\lesseqqgtr="3\msx@53
\mathchardef\gtreqqless="3\msx@54
\mathchardef\Rrightarrow="3\msx@56
\mathchardef\Lleftarrow="3\msx@57
\mathchardef\veebar="2\msx@59
\mathchardef\barwedge="2\msx@5A
\mathchardef\doublebarwedge="2\msx@5B
\mathchardef\angle="0\msx@5C
\mathchardef\measuredangle="0\msx@5D
\mathchardef\sphericalangle="0\msx@5E
\mathchardef\varpropto="3\msx@5F
\mathchardef\smallsmile="3\msx@60
\mathchardef\smallfrown="3\msx@61
\mathchardef\Subset="3\msx@62
\mathchardef\Supset="3\msx@63
\mathchardef\Cup="2\msx@64

\mathchardef\Cap="2\msx@65

\mathchardef\curlywedge="2\msx@66
\mathchardef\curlyvee="2\msx@67
\mathchardef\leftthreetimes="2\msx@68
\mathchardef\rightthreetimes="2\msx@69
\mathchardef\subseteqq="3\msx@6A
\mathchardef\supseteqq="3\msx@6B
\mathchardef\bumpeq="3\msx@6C
\mathchardef\Bumpeq="3\msx@6D
\mathchardef\lll="3\msx@6E

\mathchardef\ggg="3\msx@6F

\mathchardef\circledS="0\msx@73
\mathchardef\pitchfork="3\msx@74
\mathchardef\dotplus="2\msx@75
\mathchardef\backsim="3\msx@76
\mathchardef\backsimeq="3\msx@77
\mathchardef\complement="0\msx@7B
\mathchardef\intercal="2\msx@7C
\mathchardef\circledcirc="2\msx@7D
\mathchardef\circledast="2\msx@7E
\mathchardef\circleddash="2\msx@7F
\def\ulcorner{\delimiter"4\msx@70\msx@70 }
\def\urcorner{\delimiter"5\msx@71\msx@71 }
\def\llcorner{\delimiter"4\msx@78\msx@78 }
\def\lrcorner{\delimiter"5\msx@79\msx@79 }
\def\yen{\mathhexbox\msx@55 }
\def\checkmark{\mathhexbox\msx@58 }
\def\circledR{\mathhexbox\msx@72 }
\def\maltese{\mathhexbox\msx@7A }
\mathchardef\lvertneqq="3\msy@00
\mathchardef\gvertneqq="3\msy@01
\mathchardef\nleq="3\msy@02
\mathchardef\ngeq="3\msy@03
\mathchardef\nless="3\msy@04
\mathchardef\ngtr="3\msy@05
\mathchardef\nprec="3\msy@06
\mathchardef\nsucc="3\msy@07
\mathchardef\lneqq="3\msy@08
\mathchardef\gneqq="3\msy@09
\mathchardef\nleqslant="3\msy@0A
\mathchardef\ngeqslant="3\msy@0B
\mathchardef\lneq="3\msy@0C
\mathchardef\gneq="3\msy@0D
\mathchardef\npreceq="3\msy@0E
\mathchardef\nsucceq="3\msy@0F
\mathchardef\precnsim="3\msy@10
\mathchardef\succnsim="3\msy@11
\mathchardef\lnsim="3\msy@12
\mathchardef\gnsim="3\msy@13
\mathchardef\nleqq="3\msy@14
\mathchardef\ngeqq="3\msy@15
\mathchardef\precneqq="3\msy@16
\mathchardef\succneqq="3\msy@17
\mathchardef\precnapprox="3\msy@18
\mathchardef\succnapprox="3\msy@19
\mathchardef\lnapprox="3\msy@1A
\mathchardef\gnapprox="3\msy@1B
\mathchardef\nsim="3\msy@1C
\mathchardef\napprox="3\msy@1D
\mathchardef\nsubseteqq="3\msy@22
\mathchardef\nsupseteqq="3\msy@23
\mathchardef\subsetneqq="3\msy@24
\mathchardef\supsetneqq="3\msy@25
\mathchardef\subsetneq="3\msy@28
\mathchardef\supsetneq="3\msy@29
\mathchardef\nsubseteq="3\msy@2A
\mathchardef\nsupseteq="3\msy@2B
\mathchardef\nparallel="3\msy@2C
\mathchardef\nmid="3\msy@2D
\mathchardef\nshortmid="3\msy@2E
\mathchardef\nshortparallel="3\msy@2F
\mathchardef\nvdash="3\msy@30
\mathchardef\nVdash="3\msy@31
\mathchardef\nvDash="3\msy@32
\mathchardef\nVDash="3\msy@33
\mathchardef\ntrianglerighteq="3\msy@34
\mathchardef\ntrianglelefteq="3\msy@35
\mathchardef\ntriangleleft="3\msy@36
\mathchardef\ntriangleright="3\msy@37
\mathchardef\nleftarrow="3\msy@38
\mathchardef\nrightarrow="3\msy@39
\mathchardef\nLeftarrow="3\msy@3A
\mathchardef\nRightarrow="3\msy@3B
\mathchardef\nLeftrightarrow="3\msy@3C
\mathchardef\nleftrightarrow="3\msy@3D
\mathchardef\divideontimes="2\msy@3E
\mathchardef\varnothing="0\msy@3F
\mathchardef\nexists="0\msy@40
\mathchardef\mho="0\msy@66
\mathchardef\thorn="0\msy@67
\mathchardef\beth="0\msy@69
\mathchardef\gimel="0\msy@6A
\mathchardef\daleth="0\msy@6B
\mathchardef\lessdot="3\msy@6C
\mathchardef\gtrdot="3\msy@6D
\mathchardef\ltimes="2\msy@6E
\mathchardef\rtimes="2\msy@6F
\mathchardef\shortmid="3\msy@70
\mathchardef\shortparallel="3\msy@71
\mathchardef\smallsetminus="2\msy@72
\mathchardef\thicksim="3\msy@73
\mathchardef\thickapprox="3\msy@74
\mathchardef\approxeq="3\msy@75
\mathchardef\succapprox="3\msy@76
\mathchardef\precapprox="3\msy@77
\mathchardef\curvearrowleft="3\msy@78
\mathchardef\curvearrowright="3\msy@79
\mathchardef\digamma="0\msy@7A
\mathchardef\varkappa="0\msy@7B
\mathchardef\hslash="0\msy@7D
\mathchardef\hbar="0\msy@7E
\mathchardef\backepsilon="3\msy@7F
\def\Bbb{\ifmmode\let\next\Bbb@\else
 \def\next{\errmessage{Use \string\Bbb\space only in math mode}}\fi\next}
\def\Bbb@#1{{\Bbb@@{#1}}}
\def\Bbb@@#1{\fam\msyfam#1}

\def\inv{^{\raise.15ex\hbox{${
  \scriptscriptstyle -}$}\kern-.05em 1}}

\def\Dsl{\,\raise.15ex\hbox{$/$}\mkern-13.5mu D}
\def\dsl{\raise.15ex\hbox{$/$}\kern-.57em\hbox{$\partial$}}

\def\lspace{\ifx\answ\bigans{}\else\qquad\fi}

\def\CA{\hbox{{$\cal A$}}} \def\CC{\hbox{{$\cal C$}}}

\def\CR{\hbox{{$\cal R$}}}

\def\lform{\hbox{$\sqcup$}\llap{\hbox{$\sqcap$}}}
\def\darr#1{\raise1.5ex\hbox{$\leftrightarrow$}
\mkern-16.5mu #1}

\def\h{{{1\over2}}}

\def\INT{{\textstyle \int\kern-.642em\int}}

\def\R{{\Bbb R}}
\def\C{{\Bbb C}}
\def\Z{{\Bbb Z}}

\def\eps{{\epsilon}}

\def\Rep{{\rm Rep\, }}

\def\cocross{{>\!\!\!\triangleleft}}

\def\tens{\mathop{\otimes}}

\def\la{{\triangleright}}\def\ra{{\triangleleft}}

\def\isom{{\cong}}

\def\id{{\rm id}}

\def\Lin{{\rm Lin}}

\def\nquad{{\!\!\!\!\!\!}}
\def\nqquad{\nquad\nquad}
\def\eqn#1#2{\begin{equation}#2\label{#1}\end{equation}}

\def\o{{}_{(1)}}\def\t{{}_{(2)}}\def\th{{}_{(3)}}

\def\bo{{}^{\bar{(1)}}}\def\bt{{}^{\bar{(2)}}}

\def\und#1{{\underline {#1}}}

\def\uo{{{}^{(1)}}}\def\ut{{{}^{(2)}}}

\def\new#1{\goodbreak\goodbreak\bigskip
\noindent{\bf #1}}

\def\text#1{\mbox{\rm #1}}
\def\note#1{}

\def\blacksquare{{\lform}}
\def\frac#1#2{{{#1\over#2}}}

\def\proof{\goodbreak\noindent{\bf Proof\quad}}

\def\endproof{{\ $\lform$}\bigskip }

\def\align#1{\begin{eqnarray*}#1\end{eqnarray*}}

\def\und#1{{\underline{#1}}}

\def\Bo{{{}_{\und{(1)}}}}\def\Bt{{{}_{\und{(2)}}}}
\def\vect{{\bf t}}
\def\vecu{{\bf u}}

\def\<{\langle}
\def\>{\rangle}

\makeatother

\documentstyle[11pt]{article}

\def\thebibliography#1{\section*{REFERENCES}\list
 {[\arabic{enumi}]}{\settowidth\labelwidth{[#1]}\leftmargin\labelwidth
 \advance\leftmargin\labelsep
 \usecounter{enumi}}
 \def\newblock{\hskip .11em plus .33em minus -.07em}
 \sloppy
 \sfcode`\.=1000\relax}

\newtheorem{lemma}{Lemma}[section]
\newtheorem{propos}[lemma]{Proposition}
\newtheorem{example}[lemma]{Example}

\newtheorem{corol}[lemma]{Corollary}
\newtheorem{defin}[lemma]{Definition}

\begin{document}\baselineskip 25pt

{\ }\hskip 4.7in DAMTP/92-48
\vspace{.5in}

\begin{center} {\Large THE QUANTUM DOUBLE AS QUANTUM MECHANICS}
\baselineskip 10pt{\ }\\
{\ }\\ S. Majid\footnote{SERC Fellow and Drapers Fellow of Pembroke College,
Cambridge}\\ {\ }\\
Department of Applied Mathematics\\
\& Theoretical Physics\\ University of Cambridge\\ Cambridge CB3 9EW, U.K.
\end{center}

\begin{center}
September 1992 -- revised March 1993\end{center}
\vspace{10pt}
\begin{quote}\baselineskip 13pt
\centerline{ABSTRACT}
We introduce $*$-structures on braided groups and braided matrices. Using this,
we show that the quantum double $D(U_q(su_2))$ can be viewed as the quantum
algebra of observables of a quantum particle moving on a hyperboloid in
q-Minkowski space (a three-sphere in the Lorentz metric), and with the role of
angular momentum played by $U_q(su_2)$. This provides a new example of a
quantum system whose algebra of observables is a Hopf algebra. Furthermore, its
dual Hopf algebra can also be viewed as a quantum algebra of observables, of
another quantum system. This time the position space is a q-deformation of
$SL(2,\R)$ and the momentum group is $U_q(su_2^*)$ where $su_2^*$ is the
Drinfeld dual Lie algebra of $su_2$. Similar results hold for the quantum
double and its dual of a general quantum group.

\bigskip
\noindent {\em Keywords: quantum groups, quantum double, non-commutative
geometry, Mackey quantization, duality, Lorentz  metric}
\end{quote}
\baselineskip 20pt

\section{Introduction} One of the most important quantum group constructions is
 Drinfeld's
quantum double\cite{Dri}. It has a rich algebraic structure and, moreover,
plays an important (though not fully
understood) role in quantum inverse scattering as some kind of quantum dressing
transform\cite{JurSto:dre}. It has also been proposed as a kind of
`complexification', for example the double of $U_q(su_2)$ has been proposed as
a quantum Lorentz group\cite{PodWor:def}. In both of these contexts the quantum
double plays the physical role of a kind of generalized symmetry.

In this paper we give a new physical interpretation of the quantum double, as
the algebra of observables of a quantum mechanical system. A further
interpretation as a quantum frame bundle is given in \cite{BrzMa:gau}.
These interpretations are made possible by results about the algebraic
structure of the quantum double in \cite{Ma:dou} and \cite{Ma:skl}, to which
the present paper is a sequel. The main result of these works is that the
quantum double of a true quantum group (with universal $R$-matrix or
quasitriangular structure\cite{Dri}) has the structure of a semidirect product.
This opens up the possibility of a quantum mechanical interpretation in the
context of Mackey quantization\cite{Mac:ind}\cite{DobTol:mec} and its natural
generalization to quantum groups\cite{Ma:phy}. Of course for a quantum
mechanical interpretation, we need to extend the semidirect product result for
the quantum double to the level of $*$-algebras, and this is the main goal of
the present paper from a mathematical point of view. In doing this, we will
have to study $*$-structures on quantum groups in relation to their
quasitriangular structure $\CR$ as well as $*$-structures on certain associated
braided groups\cite{Ma:bra}\cite{Ma:bg}. This is the topic of Section~2.
We distinguish two natural possibilities, namely $\CR^{*\tens *}=\CR_{21}$
(the real case) and $\CR^{*\tens *}=\CR^{-1}$ (the antireal case). The first
possibility has been noted some time ago in a classification by Lyubashenko
\cite{Lyu:rea} and applies to the compact forms of the standard quantum
deformations $U_q(g)$. The second possibility seems to be more novel, and
applies
for example to $U_q(sl(2,\R))$. Both are needed in the paper.

The quantum mechanical picture of the double is obtained in Section~3. In fact,
we have already asked in a series of papers the following question: when is the
algebra of observables of a quantum system a Hopf algebra (and if so, what does
it mean physically)? In answer to this question we found a large class of
homogeneous spaces such that the algebras of observables of quantum mechanics
on them (via Mackey quantization) were indeed Hopf algebras. Not any
homogeneous space
satisfies this and the ones that do so arise in pairs from the factorization of
any group into two subgroups (the two factors then act on each other giving two
matching homogeneous spaces). Moreover, the dual Hopf algebra of the
quantization of one homogeneous space is the Hopf algebra of observables of
quantization of the other one. The meaning of the coproduct is that of a
non-Abelian group structure on phase-space, describing some kind of
non-commutative geometry\cite{Con:alg}. The duality means that the quantum
algebra of one of the homogeneous spaces is equivalent to the coalgebra (hence
geometry) of the
other. This possibility of a dual interpretation as quantum mechanics on the
one hand and geometry on the other was one of the main ideas introduced in the
authors PhD Thesis and we will explore it below for our new examples based on
the quantum double. In purely quantum mechanical terms it means that the states
of the system also form an algebra and hence a certain symmetry is restored
between observables and states. For further details we refer to
\cite{Ma:phy}\cite{Ma:mat}\cite{Ma:hop}\cite{Ma:pla}\cite{Ma:pri}.

The self-duality of the situation is also connected with other dualities in
physics and also with mirror symmetry in string theory. In particular, a
natural source of factorizations is provided by the Iwasawa decomposition
$G_\C=G^*G$ of the complexification of compact semisimple Lie group $G$ into
$G$ and a solvable group $G^*$. The resulting action of $G$ on $G^*$ gives one
system with $G^*$ position and $G$ momentum, while at the action of $G^*$ on
$G$ which arises in the same process gives the dual system with the roles of
the position and momentum interchanged. Moreover, the group $G^*$ here is
connected at the Lie algebra level with considerations of Manin-triples and the
classical double in the theory of classical inverse
scattering\cite{Dri}\cite{Sem:wha}. In \cite{Ma:mat} we constructed the mutual
group actions using the holonomy of a pair of zero-curvature connections. For a
concrete example, the Iwasawa decomposition of $SL(2,\C)$ gave rise to a system
with position space given by non-concentrically nested spheres, and momentum
given by $su(2)$ (angular momentum). The spheres here are orbits under an
action of $SU(2)$ or $SO(3)$ on $\R^3$, but an unusual feature was that in
order for the quantum algebra of observables to be a Hopf algebra, this action
was not the usual rotation but a distorted non-linear one (coming from the
Iwasawa decomposition). The non-linearity moreover led to an `event-horizon'
type structure at the plane $z=-1$ in $\R^3$. In this example $SU(2)^*$ is a
solvable group which can be identified with the region $z>-1$. Such `event
horizons' appear to be a characteristic feature of the allowed homogeneous
spaces in \cite{Ma:phy}\cite{Ma:mat}\cite{Ma:hop}\cite{Ma:pla}. For example, in
1+1 dimensions the requirement of self-duality forced the metric to be a
black-hole type one.

We are going to use much the same ingredients now in our interpretation of the
quantum double and its dual. This time the actions will be more standard
rotations (in Minkowski spacetime for example) but the novel ingredient now is
that the underlying classical system can be a q-deformed geometry rather than
an ordinary classical geometry as in the models above.  Our constructions are
quite general but we concentrate on the doubles $D(U_q(g))$ of the
Drinfeld-Jimbo quantum groups $U_q(g)$\cite{Dri}\cite{Jim:dif}, giving all
formulas explicitly for the case where $g=su_2$.

In this case we take for our q-deformed position observables the $*$-algebra
$BS^3_q$ consisting of $BH_q(2)$ (the $*$-algebra of $2\times 2$ hermitian
braided-matrices) modulo the condition $BDET=1$ where $BDET$ is the
braided-determinant. The algebra of $2\times 2$ braided-matrices has been
introduced in\cite{Ma:exa} and has four generators $\pmatrix{a&b\cr c&d}$ to be
regarded as the `co-ordinate functions' on the braided space. We equip this now
with the $*$-structure $\pmatrix{a^*&b^*\cr c^*& d^*}=\pmatrix{a&c\cr b&d}$
appropriate to the hermitian case, and impose the further determinant
condition. Recall that in the undeformed case the Hermitian matrices can be
identified with Minkowski space and the determinant with the Lorentz metric.
The algebra $BS^3_q$ is therefore a q-deformation of the co-ordinate functions
on the sphere $S^3_{\rm Lor}$ of unit radius in Minkowski space. Meanwhile, the
q-deformed momentum of our system is given by the quantum group $U_q(su_2)$.
Recall that in the undeformed situation $SL(2,\C)$ acts on the hermitian
matrices $H(2)$ (Minkowski space) by conjugation and preserves the determinant.
Its subgroup $SU(2)$ acts precisely by spatial rotations of Minkowski space and
of $S^3_{\em Lor}$. In our q-deformed setting the action of $U_q(su_2)$
corresponds precisely to this action by conjugation, in the form of the quantum
adjoint action.

By quantization of this classical q-deformed system we mean a $*$-algebra
containing the position observables and momentum (quantum) group in such a way
that the action of momentum is implemented in the algebra. This we understand
as the construction of a (quantum group) $*$-algebra cross product which is a
natural generalization of the Mackey quantization scheme used
above\cite{Ma:phy}. A recent work in which  $*$-algebra cross products are
understood as quantization is in \cite{Egu:mec}. In our case the above
q-deformed classical system has such a quantization. Moreover, it is a Hopf
algebra and is isomorphic as a Hopf algebra to Drinfelds quantum double
$D(U_q(su_2))$. Full details of this example appear at the end of Section~3
below.

Note that $q$ here is considered quite orthogonal to the process of
quantization, and need not be related to any physical $\hbar$. Thus in our
interpretation we have two independent processes, which mutually commute,
\eqn{double.int}{\matrix{ C(S^3_{\rm Lor})\tens su_2&{\buildrel {\rm
deformation}\over \to}& BS^3_q\tens U_q(su_2)\cr
& & \cr
{\rm \scriptstyle quantization}\ \downarrow& &\downarrow{\rm \scriptstyle \
quantization}\cr
& & \cr
C(S^3_{\rm Lor})\cocross U(su_2)&{\buildrel {\rm deformation}\over
\to}&BS^3_q\cocross U_q(su_2).}}
The top left here can be viewed as a subset of the  classical observables
$C(T^*S^3_{\rm Lor})$ consisting of certain functions on $T^*S^3_{\rm Lor}$
that are linear in the fiber direction, namely the tensor product of functions
on the base $S^3_{\rm Lor}$ and vector vector fields induced by the action of
$su_2$. This is the natural subset that is quantized in the Mackey scheme (cf
the vertical polarization). The bottom left is the usual Makey quantization of
this system and is isomorphic as an algebra to Drinfelds double $D(SU(2))$. We
have already explained in \cite[Example~2.4]{Ma:phy} that Drinfeld's quantum
double of a group or enveloping algebra is a semidirect product and hence forms
an example of the general class of models on homogeneous spaces as discussed
above. The rest of the diagram is filled in by the constructions in the present
paper. Note in this context that the classical situation works just as well
with position observables $SU(2)=S^3$ (in Euclidean space). The action of
momentum $SU(2)$ is again by conjugation, with orbits again spheres. The
quantization as an algebra is again Drinfeld's quantum double $C(SU(2))\cocross
U(su_2)\isom D(SU(2))$. The only difference is that the $*$-structure on the
matrix co-ordinate functions is that appropriate to unitary rather than to
Hermitian matrices as above. On the other hand the naive q-deformation of this
Euclidean system to $SU_q(2)\cocross U_q(su_2)$ is not possible because the
quantum adjoint action fails to respect the algebra structure of $SU_q(2)$ for
$q\ne 1$. In fact, I do not know how to q-deform this Euclidean situation and
have been forced by this failure into the Minkowski setting and its
q-deformation via braided matrices rather than more familiar quantum ones.

One can ask why should we be interested in a quantum system whose underlying
classical system is more naturally a $q$-deformed classical geometry? One
reason is that we have an extra $q$ parameter to regularise any singularities
that appear in the quantum theory (even if, after renormalization, we set
$q=1$)\cite{Ma:reg}. Related to this it is interesting that the requirement of
existence of a q-deformation leads us into the Minkowski setting even if we are
interested in the end only in $q=1$. Another motivation is that at the Planck
scale we can expect some feedback between
quantization and geometry in the sense that a correct formulation of quantum
particles may surely require them to be moving in the background of something
other than a usual geometry. Replacing the latter by a $q$-geometry is one
possibility for such models, with $q$ expressing quantum corrections to the
background geometry itself.

In general, while the need for some kind of non-commutative or $q$-geometry has
become clear, there remains a shortage of natural examples and physical
principles to govern such a geometry (simply asking to $q$-deform everything
still leaves a lot of possibilities). Hence it is significant that the quantum
double provides a natural example, the study of which can help develop the
subject further. For example, one can study $q$-deformed differential
structures and hope to give a more conventional (but $q$-deformed) picture of
the quantum double as quantizing a $q$-symplectic or $q$-Poisson space. One can
also try to take a different line and combine the two steps in
(\ref{double.int}) into a single quantization of a pair of compatible Poisson
brackets as in \cite{DGM:mat}. We will not attempt these steps here, but see
the concluding remarks at the end of the paper.

Let us recall now that our original motivation for quantum systems whose
algebra of observables are Hopf algebras was an interesting quantum/gravity
duality phenomenon implemented by Hopf algebra duality. This is the topic of
Section~4 where we study the dual of the quantum double. It is also a Hopf
algebra, but our result is that it too is a cross product quantization. Here
again we depend on the general algebraic theory in \cite{Ma:skl} but in the
context now of $*$-algebras and in more suitable right-handed conventions. The
most unusual feature of the result is that this time, for the dual of the
quantum double to be a $*$-algebra cross product (as needed for its
interpretation as generalized Mackey quantization), the quasitriangular
structure should be antireal rather than real. For example, it is the dual of
the quantum double of $U_q(sl(2,\R))$ that has the desired interpretation,
rather than of the compact real form $U_q(su_2)$ above. Thus we have for the
dual of the quantum double in this case the interpretation
\eqn{codouble.int}{ \matrix{C(SL(2,\R))\tens su_2^*&{\buildrel {\rm
deformation}\over \to}& BSL_q(2,\R)\tens U_q(su_2^*)\cr
& & \cr
{\rm \scriptstyle quantization}\ \downarrow& &\downarrow{\rm \scriptstyle \
quantization}\cr
& & \cr
C(SL(2,\R))\cocross U(su_2^*)&{\buildrel {\rm deformation}\over \to}&
BSL_q(2,\R)\cocross U_q(su_2^*).}}
The role of angular momentum is now played by the Drinfeld dual $su_2^*$ as
mentioned above. Its $q$-deformation is by definition
$U_q(su_2^*)=U_q(sl_2)^*$, i.e., the quantum-group function algebra dual to
$U_q(sl_2)$ but regarded nevertheless as a quantum enveloping algebra. The role
of position is again played in the deformed case by the braided group
$BSL_q(2,\R)$ which is like the above but with $*$-structure
$\pmatrix{a^*&b^*\cr c^*& d^*}=\pmatrix{a&b\cr c&d}$. Full details appear at
the end of Section~4. The Mackey-type quantization of this system is then the
dual Hopf algebra to the quantum double and hence dual to the system
(\ref{double.int}). This possibility of a dual interpretation is an interesting
direction for further work and suggests a genuinely new physical phenomenon.
According to the duality principle introduced in \cite{Ma:pla}, this dual
system should appear relative to our initial system as physics `beyond' the
Planck scale. This is necessarily a speculative topic but its elaboration
remains a long-term motivation for the present work.

We conclude in Section~5 with some remarks connecting the constructions here
with other points of view in \cite{Ma:bos}\cite{BrzMa:gau}. We explain in the
last of these that the quantum algebras of observables above can just as easily
be understood in a geometric way as principal bundles on quantum homogeneous
spaces. When the theory of principal bundles is formulated in the setting of
non-commutative algebraic geometry there is not really any difference between
semidirect products viewed as quantization and semidirect products viewed as
(algebraic) geometry. Hence the double and its dual both have this geometrical
interpretation as well as the quantum one above. This is an important principle
that is surely relevant to Planck-scale physics where geometry and quantum
theory need to be unified.

Throughout this paper we work over a field with involution, which we fix for
concreteness to be $\C$. The general results hold for an arbitrary field with
involution. Thus our approach to quantization is an algebraic one as explained
in detail in \cite[Sec. 1.1.1]{Ma:phy}. The geometrical content is also to be
understood in a setting of non-commutative algebraic geometry as indicated
above. One can try to place these results explicitly in a $C^*$-algebra or von
Neumann algebra setting though we shall not attempt to do so here.  This, and
the interpretation of the $q$-deformed Mackey construction in terms of
$q$-deformed symplectic structures etc are two directions for further work. The
present paper is a necessary first step.

\new{Preliminaries}

Here we collect some basic algebraic facts about Hopf algebra cross products
and cross coproducts, the quantum double and braided groups.

Recall first that a Hopf algebra is $(H,\Delta,\eps,S)$ where $H$ is an algebra
with unit, $\Delta:H\to H\tens H$ (the comultiplication), $\eps:H\to \C$ (the
counit) are algebra maps and $S:H\to H$ plays the role of inverse. An
introduction to the axioms is in \cite{Ma:qua}. We often use the formal sum
notation $\Delta h=\sum h\o\tens h\t$ \cite{Swe:hop}. If $H$ is finite
dimensional then $A=H^*$ is also a Hopf algebra, dual to $H$, with
multiplication determined by the comultiplication in $H$ and vice-versa. A
similar situation holds in the infinite-dimensional case if we work with a
$q$-adic or other topology as in \cite{Dri}. An algebraic alternative, which we
adopt, is simply to work with dually paired Hopf algebras. Thus $U_q(su_2)$ is
nondegenerately paired with $SU_q(2)$ etc, in a standard way. For brevity some
of our abstract results are proven in the finite-dimensional case: their
extension to the dually-paired case is straightforward and verified directly
when we come to the relevant examples.

When a Hopf algebra $H$ acts on an algebra $B$ in such a way as to respect its
structure in the sense
\eqn{p1}{ h\la (bc)=\sum (h\o\la b)(h\t\la c),\qquad h\la 1=\eps(h)1}
(one says that $B$ is an $H$-module algebra), one may form a {\em cross
product} algebra $B\cocross H$. This is just as familiar for group actions. It
is built on $B\tens H$ with product
\eqn{p2}{(b\tens h)(c\tens g)=\sum b(h\o\la c)\tens h\t g,\qquad b,c\in B,\
h,g\in H.}
Put another way, $B\cocross H$ has $B,H$ as subalgebras and cross relations
given by $(1\tens h)(b\tens 1)=\sum (h\o\la b\tens 1)(1\tens h\t)$. Also, a
feature of Hopf algebras is that Hopf algebra constructions also have dual
versions obtained by writing the relevant maps as arrows and reversing them.
Thus, if $B$ is a coalgebra on which $H$ coacts by $\beta:B\to H\tens B$ in
such a way as to preserve its structure,
\eqn{p4}{(\id\tens\Delta_B)\beta (b)=\sum b\o\bo b\t\bo\tens b\o\bt\tens
b\t\bt,\qquad (\id\tens\eps_B)\beta(b)=1\eps_B(b)}
where $\beta(b)=\sum b\bo\tens b\bt$ (one says that $B$ is an $H$-comodule
coalgebra), we can form a {\em cross coproduct coalgebra} $B\cocross H$.
Explicitly, it is built on $B\tens H$ with coproduct
\eqn{p3}{\Delta(b\tens h)=\sum b\o\tens b\t\bo h\o\tens b\t\bt\tens h\t.}
An introduction to these topics is in \cite[Sec. 6]{Ma:qua}.

If $H$ is a (say, finite-dimensional) Hopf algebra there is a {\em quantum
double} Hopf algebra $D(H)$ built on $H^*\tens H$ as follows. The
comultiplication, counit and unit are the tensor product ones. So
$\Delta(a\tens h)=\sum a\o\tens h\o\tens a\t\tens h\t$.
The product is twisted
\eqn{p5}{ (a\tens h)(b\tens g)=\sum  b\t a\tens h\t g
<Sh\o,b\o><h\th,b\th>,\qquad h,g\in H;\ a,b\in A=H^*.}
This Hopf algebra was introduced by
Drinfeld via generators and relations\cite{Dri}. We use here the abstract form
due to the author in \cite{Ma:phy} and in the conventions of \cite{Ma:dou}.
Note that $H$ and $H^*{}^{\rm op}$ ($H^*$ with the opposite product) are
sub-Hopf algebras. Moreover, the quantum double factorizes into these two
factors. The general theory of factorizations of Hopf algebras was introduced
in \cite{Ma:phy}. One shows that for any such factorization, the two factors
act on each other and the entire Hopf algebra can be recovered from the factors
as a double-cross product. Thus $D(H)\isom H\bowtie H^{*\rm op}$ where the
actions are coadjoint actions\cite[Example 4.6]{Ma:phy}. In the
infinite-dimensional case we use a formulation in terms of dually paired Hopf
algebras along the lines given in \cite{Ma:seq}. Also, the dual $D(H)^*$ of the
quantum double is also a Hopf algebra. It is built on $H\tens H^*$ as an
algebra, with a twisted comultiplication
\eqn{p6}{\Delta (h\tens a)=\sum h\t\tens (Sf^b\o)a\o f^b\th\tens e_b\tens
a\t<f^b\t,h\o>}
where $\{e_b\}$ is a basis of $H$ and $\{f^b\}$ a dual basis.

A Hopf algebra $H$ is quasitriangular (a quantum group of enveloping algebra
type) if there is an invertible element $\CR\in H\tens H$ obeying the axioms of
Drinfeld\cite{Dri},
\eqn{p7}{(\Delta\tens\id)\CR=\CR_{13}\CR_{23},\quad
(\id\tens\Delta)\CR=\CR_{13}\CR_{12},\quad \Delta^{\rm op}=\CR(\Delta\
)\CR^{-1}}
where $\Delta^{\rm op}$ denotes the opposite comultiplication. We define
$Q=\CR_{21}\CR_{12}$ to be the {\em quantum Killing form}. A quasitriangular
Hopf algebra $H$ is called factorizable\cite{ResSem:mat} if $Q$ is
non-degenerate in the sense that the set $\{(\phi\tens\id)(Q)|\phi\in
H^*\}\subseteq H$ coincides with all of $H$. The quantum double of any Hopf
algebra is quasitriangular and indeed, factorizable, as are many familiar
quantum groups such as $U_q(g)$ (working with suitable topological completions
or with a designated dually-paired Hopf algebra in the role of $H^*$). Clearly,
the dual concept of a quasitriangular Hopf algebra is a Hopf algebra $A$
equipped with a map $\CR:A\tens A\to \C$ obeying some obvious axioms dual to
(\ref{p7}), namely
\eqn{p8}{\CR(ab\tens c)=\sum
\CR(a\tens c\o)\CR(b\tens c\t),\quad \CR(a\tens bc)=\sum
\CR(a\o\tens c)\CR(a\t\tens b)}
\eqn{p9}{ \sum b\o a\o \CR(a\t\tens b\t)=\sum \CR(a\o\tens b\o) a\t b\t}
for all $a,b,c\in A$.  The dual quantum groups (quantum groups of function
algebra type) are like this. Finally, the axioms of a (dual) quasitriangular
Hopf algebra are such that its category of (co)modules forms a braided or
quasitensor category. If $V,W$ are $H$-modules, the  braided-transposition or
braiding is
\eqn{p10}{ \Psi_{V,W}:V\tens W\to W\tens V,\quad \Psi_{V,W}(v\tens
w)=\tau(\CR\la(v\tens w))}
where $\CR\in H\tens H$ acts on $V\tens W$. It is an intertwiner for the action
of $H$ and has properties similar to those of the usual transposition $\tau$,
except that $\Psi_{V,W}$ and $\Psi_{W,V}^{-1}$ need not coincide (they are
usually represented as distinct braids). Similarly for the category of
comodules in the dual quasitriangular case.

Next we need the notion of an algebra {\em living in} a braided category. We
will be concerned here only with the braided categories of representations of a
quantum group as just explained. In this case, an algebra in the category means
an ordinary algebra $B$, on which the quantum group acts, in such a way that
the algebra and unit maps are covariant. The tensor product action on $B\tens
B$ is given by the action of $\Delta(H)$. Hence the necessary condition is just
that $B$ is an $H$-module algebra as in (\ref{p1}). We saw above that this led
to cross products. On the other hand our new category-theoretical view of
$H$-module algebras is also very useful and leads to the notion of {braided
tensor products} of $H$-module algebras when $H$ is a quantum
group\cite{Ma:bra}. The idea is that if $B,C$ are two algebras living in a
braided category, then $B\tens C$ has an algebra structure, which we denote
$B\und\tens C$, also living in the category. In our present case this is
\eqn{p12}{ (b\tens c)(d\tens e)=b\Psi_{C,B}(c\tens d)e=\sum b(\CR\ut\la d)\tens
(\CR\uo\la c)e}
where the second expression uses (\ref{p10}). One can check that this is
associative.

This is one of the fundamental constructions behind the theory of braided
groups. For a braided group $B$ is a kind of Hopf algebra in which the
braided-coproduct is taken as an algebra homomorphism $\und\Delta:B\to
B\und\tens B$ with the braided tensor product structure $B\und\tens B$ from
(\ref{p12}). The structure $B=\und H$ needed above is an example. Explicitly,
\eqn{p13}{\und H=\cases{H&as\ algebra\cr \und\Delta &modified\ comultiplication
};\quad \und\Delta b=\sum b\o S\CR\ut\tens \CR\uo\la b\t}
where $\la$ is the quantum adjoint action $h\la b=\sum h\o b S h\t$. There is
also a modified antipode. Thus $\und H$ is not an ordinary Hopf algebra but
instead, $\und\Delta:\und H\to \und H\und\tens \und H$ is an algebra
homomorphism provided $\und H$ is treated with braid statistics as in
(\ref{p12}). It is the braided group associated to $H$ (of enveloping algebra
type) and lives in the braided category of $H$-modules by the quantum adjoint
action.

There is also a braided group of function algebra
type associated to a Hopf algebra $A$ dual to $H$. It is
\eqn{p14}{ \und A=\cases{A&as\ coalgebra\cr \und\cdot&modified product};\quad
a\und\cdot b=\sum a\t b\t\CR((S a\o)a\th\tens Sb\o).}
It also has a modified antipode. This $\und A$ lives in the braided category of
left $H$-modules (or right $A$-comodules) by the coadjoint action
\eqn{p15}{h\la a=\sum <h,a\t>(Sa\o)a\th;\quad a\in A,\ h\in H}
(or right adjoint coaction $\beta(a)=\sum a\t\tens (Sa\o)a\th$). When $H$ is
factorizable and has dual $A$, then one finds remarkably that $\und H\isom\und
A$ as braided groups\cite{Ma:skl}. For a systematic introduction to braided
groups one can see \cite{Ma:introp}.

Armed with these various ingredient we have shown in \cite{Ma:skl}
cf\cite{Ma:dou} that
\eqn{double.semi}{ D(H)\isom \und A\cocross H}
when $H$ is quasitriangular and $\und A$ is associated braided group of
function algebra type. The motivation in \cite{Ma:skl} was in relation to the
role of $D(H)$ as a complexification of $H$. Our new goal in the present paper
is to give the interpretation of this semidirect product theorem as quantum
mechanics on the `braided-space' underlying $\und A$.

\section{$*$-Structures on Braided Groups}

If we are to develop a quantum-mechanical interpretation of the quantum double,
we are going to have to work with $*$-algebras. The $*$-algebra structure on a
quantum algebra of observables determines such things as positivity of states,
the possibility of Hilbert space representations etc. Such $*$-structures on
Hopf algebras are well-known since the pioneering work of \cite{Wor:twi}, but
the theory of $*$-structures on braided groups (which we will need in later
sections) has not previously been developed. This is our goal in the present
section.  After this, one may construct norms and obtain operator versions of
the various Hopf algebras and braided groups, along established lines such as
in \cite{Wor:twi}.

Let us recall that a Hopf $*$-algebra is a Hopf algebra where $H$ is itself a
$*$-algebra and $\Delta,\eps$ are $*$-algebra homomorphisms. In addition, one
requires $(S\circ *)^2=\id$\cite{Wor:com}. In this case the dual, $A$, is also
a $*$-algebra. Its $*$-structure is related to that of $H$ via
\eqn{dualstar}{ <h^*,a>=\overline{<h, (Sa)^*>},\qquad \forall h\in H,\ a\in A.}
Now, when $H$ is a $*$-quantum group (a quasitriangular Hopf $*$-algebra) it is
 natural to impose either of the two conditions
\eqn{reality}{\CR^{*\tens *}=\CR_{21}\ {\rm (real)},\qquad \CR^{*\tens
*}=\CR^{-1}\ {\rm (antireal)}.}
This is because $H^{\rm op}$ has two natural quasitriangular structures, namely
 $\CR_{21}$ and $\CR^{-1}$ and so one would expect that $*:H\to H^{\rm op}$
should map $\CR$ to one or other of them. The first of these cases has been
noted already in \cite{Lyu:rea} and as well as in \cite{Egu:mec} in the context
of quantum mechanics on the quantum sphere.

\begin{example} Let $H$ be a Hopf $*$-algebra, then so is $D(H)$ with $(a\tens
1)^*=(S^2a)^*\tens 1$, $(1\tens h)^*=(1\tens h^*)$. Its canonical
quasitriangular structure is antireal.
\end{example}
\proof For simplicity we consider only the finite-dimensional case, but see
\cite{Ma:seq}. We first make $H^{*\rm op}$ (with the reversed product of $H^*$)
into a Hopf $*$-algebra with $*^{\rm op}=S^{-2}\circ*=*\circ S^2$ and antipode
$S^{-1}$ (any even power in place of $S^2$ will do here for $*^{\rm op}$, but
this is the one that we need in our present conventions for the quantum
double). Since the quantum double is generated by these Hopf algebras there is
then a unique possibility for its $*$-structure such that the inclusions are
$*$-algebra maps as stated. On a general element we take $(a\tens h)^*=((a\tens
1)(1\tens h))^*=(1\tens h^*)((S^2a)^*\tens 1)$. Putting this into (\ref{p5})
and using (\ref{dualstar}) we see that this is indeed a $*$-algebra structure
on $D(H)$:
\align{&&\nqquad[(a\tens h)(b\tens g)]^*=\sum (1\tens g^*)(1\tens
h\t{}^*)((S^2b\t){}^*\tens 1)((S^2a)^*\tens
1)\overline{<Sh\o,b\o><h\th,b\th>}\\
&&=\sum (1\tens g^*)(1\tens h\t{}^*)((S^2b\t){}^*\tens 1)((S^2a)^*\tens
1)<(Sh\o)^*,(Sb\o)^*><h\th{}^*,(Sb\th){}^*>\\
&&=\sum (1\tens g^*)(1\tens h^*\t)((S^2b)^*\t\tens 1)((S^2a)^*\tens
1)<h^*\o,(S^2b)^*\o><h^*\th,S(S^2b)^*\th>\\
&&=\sum (1\tens g^*)((S^2b)^*\t\t\tens h^*\t\t)((S^2a)^*\tens
1)<Sh^*\t\o,(S^2b)^*\t\o>\\
&&\qquad\qquad\qquad\qquad\qquad
\qquad<h^*\t\th,(S^2b)^*\t\th><h^*\o,(S^2b)^*\o><h^*\th,S(S^2b)^*\th>\\
&&=(1\tens g^*)((S^2b)^*\tens h^*)((S^2a)^*\tens 1)=(1\tens g^*)((S^2b)^*\tens
1)(1\tens h^*)((S^2a)^*\tens 1)=(b\tens g)^*(a\tens h)^*.}
For the fifth equality we used the duality between $H^*$ and $H$ and the
antipode axioms. Moreover, $(S\circ *)(a\tens h)=S[(1\tens h^*)((S^2a)^*\tens
1)]=S^{-3}(a^*)\tens S(h^*)$ so that $(S\circ *)^2=\id$ on $D(H)$. Since the
coalgebra of the quantum double is the tensor product one, it it equally clear
that it commutes with $*$ as it should.

The standard quasitriangular structure\cite{Dri} in our conventions is
$\CR=\sum_a (f^a\tens 1)\tens (1\tens e_a)$ where $\{e_a\}$ is a basis of $H$
and $\{f^a\}$ a dual basis. Hence $\CR^{*\tens *}=\sum_a (S^{-2}(f^a{}^*)\tens
1)\tens (1\tens e_a{}^*)=\sum (S^{-1}f'{}^a\tens 1)\tens (1\tens
e'_a)=(S\tens\id)\CR=\CR^{-1}$, where $e_a'=e_a{}^*$ is a new basis with dual
basis $f'{}^a=S^{-1}(f^a{}^*)$ according to (\ref{dualstar}).
\endproof

\begin{example} Let $H=U_q(su_2)$ equipped with its usual $*$-structure
$H^*=H$, $X_\pm^*=X_\mp$ in conventions where $[X_+,X_-]={q^{H}-q^{-H}\over
q-q^{-1}}$ with $q=e^{t\over 2}$ real.
Its standard quasitriangular structure over $\C[[t]]$ is real.
\end{example}
\proof  This fact has already been observed in \cite{Lyu:rea}\cite{Egu:mec} but
we include the proof here for completeness. The standard expression for $\CR$
for $U_q(su_2)$ is\cite{KirRes:rep}
\[ \CR=q^{H\tens H\over 2}e_q^{(1-q^{-2})q^{H\over 2}X_+\tens q^{-{H\over
2}}X_-}=e_q^{(1-q^{-2})X_+q^{-{H\over 2}}\tens X_-q^{H\over 2}}q^{H\tens H\over
2}\]
where $e_q$ is a $q$-exponential $e_q^x=\sum_{m=0}^\infty {x^m\over [[m]]!}$
and $[[m]]={1-q^{-2m}\over 1-q^{-2}}$. The second expression coincides using
the relations in the algebra. Starting with the first expression for $\CR$ we
then compute
\[\CR^{*\tens *}=e_{q^*}^{(1-q^{-2})^*(q^{H\over 2} X_+)^*\tens (q^{-{H\over
2}} X_-)^*}(q^{H\tens H\over2})^*=e_{q}^{(1-q^{-2})X_-q^{H\over 2}\tens
X_+q^{-{H\over 2}} }q^{H\tens H\over2}=\CR_{21}.\]
\endproof

\begin{example} Let $H=U_q(sl(2,\R))$  with its usual $*$-structure $H^*=-H$,
$X_\pm^*=-X_\pm$ and with $|q|=1$. Its standard quasitriangular structure over
$\C[[t]]$ is antireal.
\end{example}
\proof We compute
\align{\CR^{*\tens *}&&=e_{q^*}^{(1-q^{-2})^*(q^{H\over 2} X_+)^*\tens
(q^{-{H\over 2}} X_-)^*}(q^{H\tens
H\over2})^*=e_{q^{-1}}^{(1-q^{2})X_+q^{H\over 2}\tens X_-q^{-{H\over
2}}}q^{-{H\tens H\over2}}\\
&&=e_{q^{-1}}^{-(1-q^{-2})q^{H\over 2} X_+\tens q^{-{H\over 2}} X_-}q^{-{H\tens
H\over2}}=\left(e_{q}^{(1-q^{-2})q^{H\over 2} X_+\tens q^{-{H\over 2}}
X_-}\right)^{-1}q^{-{H\tens H\over2}}=\CR^{-1}}
by the relations in the algebra (to organise the exponent of $e_{q^{-1}}$) and
the observation $(e_q^{\lambda x})^{-1}=e_{q^{-1}}^{-\lambda x}$ with the
operator $q^{H\over 2} X_+\tens q^{-{H\over 2}} X_-$ formally in the role of
$x$. Another proof is to compute $\CR^{-1}=(S\tens\id)(\CR)$ directly from the
formula for $\CR$ and compare with the third expression for $\CR^{*\tens *}$
above. \endproof

So both real and antireal quasitriangular Hopf $*$-algebras exist (in
abundance). We are interested mainly in the real case for the following
purpose:

\begin{defin} Let $H$ be a real-quasitriangular Hopf $*$-algebra and $\CC$ its
braided category of representations. A {\em real  Hopf} $\und *$-algebra $B$
living in this category is a Hopf algebra in the category (so $\und\Delta: B\to
B\und\tens B$ to the braided tensor product algebra) such that

(a) $\und{*}:B\to B$ is a $*$-algebra structure on $B$ (so it is antilinear,
$(\und*)^2=\id$ and $(bc)^{\und*}=c^{\und*}b^{\und*}$).

(b) $\und\Delta \circ {\und*}=\tau\circ(\und{*}\tens\und{*})\circ\und\Delta$
where $\tau$ is usual transposition, and $\und\eps\circ *={}^{\overline{\
}}\und\eps$.

(c) $\und S\circ\und *= \und*\circ\und S$ where $\und S$ is the
braided-antipode.

(d) The action of $H$ on $B$ obeys $(h\la b)^{\und *}=(Sh)^*\la b^{\und *}$ for
$h\in H$, $b\in B$.
\end{defin}

Although the most general axiom system for $*$-structures on Hopf algebras in
braided categories is not known, the notion introduced here is relevant for our
quantum-mechanical purposes and we will see next that it does hold for the
braided groups $\und H,\und A$ in the Preliminaries. Both the product and
coproduct are skew with respect to $\und *$. One can easily see that given
(a),(d) in Definition~2.2 the braided tensor product algebra $B\und\tens B$ has
the $*$-algebra structure cf\cite{Egu:mec}
\[{\und{*}}_{B\und\tens B}=\tau\circ(\und*\tens\und*)\]
when $\CR$ is real. In these terms the relevant part of condition (b) is
equivalent to

(b)$'$ $\und\Delta\circ\und*={\und*}_{B\und\tens B}\circ\und\Delta$.

\begin{propos} If $H$ be a real-quasitriangular Hopf $*$-algebra then the
braided group $B=\und H$ with $\und*=*$ is a real Hopf $*$-algebra in the
braided category of $H$-modules.
\end{propos}
\proof The quantum adjoint action of $H$ on $\und H$ (which we identify with
$H$ as a $*$-algebra) always obeys condition (d) for any Hopf algebra $H$
since,
\align{ (h\la b)^*&&=\sum (h\o b S h\t)^*=\sum (Sh\t)^* b^* h\o{}^*=\sum
(S^{-1} h^*\t) b^* h^*\o\\
&&=\sum (S^{-1} h^*\t) b^* S  S^{-1}h^*\o=\sum (Sh)^*\o b^* S
(Sh)^*\t=(Sh)^*\la b^*.}
For condition (b) we compute
\align{&&\nqquad(*\tens *)\und\Delta b=\sum (b\o S\CR\ut)^*\tens (\CR\uo\la
b\t)^*\\
&&=\sum (S\CR\ut)^* b\o{}^*\tens (S\CR\uo)^*\la b\t{}^*=\sum \CR\uo b^*\o\tens
\CR\ut\la b^*\t\\
&&=\sum \CR\uo b^*\o\tens \CR\ut\o b^*\t S \CR\ut\t=\sum \CR\uo\CR'\uo
b^*\o\tens \CR'\ut b^*\t S\CR\ut\\
&&=\sum \CR\uo b^*\t S\CR'\uo\tens b^*\o  S\CR'\ut S\CR\ut=\sum \CR\uo\o b^*\t
S\CR\uo\t \tens b^*\o S\CR\ut=\tau\circ\und\Delta b^*.}
Here $\CR'$ is a second copy of $\CR$ and we used the axioms (\ref{p7}) freely
as well as $(S\tens S)(\CR)=\CR$.
The braided-counit $\und\eps$ coincides with the usual one, so like the algebra
structure, it automatically has the required properties. For the
braided-antipode we compute from the expression \cite{Ma:bra} $\und S b=\sum
\CR\ut u^{-1} S(\CR\uo b)$ where $u^{-1}=\sum \CR\ut S^2\CR\uo$,
\align{*\circ\und S b&&=\sum (S^{-1}(b^*\CR\uo{}^*)) u^{-1}\CR\ut{}^*=\sum
(S^{-1}(b^*\CR\ut))u^{-1}\CR\uo\\
&&=\sum (S^{-1}\CR\ut)u^{-1}(Sb^*)S S^{-1}\CR\uo =\sum \CR\ut u^{-1} S(\CR\uo
b^*)=\und S\circ * b.}
Here $u^{-1}{}^*=\sum (S^{-2}\CR\uo^*)\CR\ut{}^*=S^{-2}(u^{-1})=u^{-1}$ when
$\CR$ is real. \endproof

\begin{propos} Let $A$ be the dual of a real-quasitriangular Hopf $*$-algebra
$H$. The braided group $B=\und A$ with $\und *=*\circ S$ is a real Hopf
$*$-algebra in the braided category of $H$-modules.
Similarly when $A$ is dual quasitriangular and $\und A$ lives in the braided
category of $A$-comodules. \end{propos}
\proof First we verify that $\und A$ as an $H$-module under (\ref{p15}) obeys
the required condition (d) (for any Hopf algebra $H$ with dual $A$). Indeed,
\align{(h\la a)^{\und *}\nqquad&&=*\circ S\sum a\t <h,(Sa\o)S^{-1} S a\th>=\sum
(Sa)\t{}^*\overline{<h,(Sa)\th S^{-1} (S a)\o>}\\
&&=\sum (Sa)\t{}^* <(Sh)^*, (S (Sa)\o{}^*)(Sa)\th{}^*>=\sum
a^{\und*}\t<(Sh)^*,(Sa^{\und*}\o) a^{\und*}\th>=(Sh)^*\la a^{\und*}.}
Next, the quasitriangular structure on $H$ induces a dual quasitriangular
structure $\CR(a\tens b)=<\CR,a\tens b>$. From (\ref{dualstar}) and $(S\tens
S)(\CR)=\CR$, one finds that the reality condition in these dual terms is
\eqn{dual-real}{\CR(a^*\tens b^*)=\overline{\CR(b\tens a)}.}
The structure of $\und A$ was recalled in terms of this $\CR$ in the
Preliminaries. We have
\align{&&\nqquad(a\und\cdot b)^{\und *}=*\circ S(a\und \cdot b)=\sum (Sa\t)^*
(Sb\t)^*\overline{\CR((Sa\o)a\th\tens Sb\o)}\\
&&=\sum a\t{}^{\und*} b\t{}^{\und*}\CR((Sb\o)^*\tens (S(Sa\th)^*)(Sa\o)^*)=\sum
a\t{}^{\und*} b\t{}^{\und*}\CR(b\o{}^{\und*}\tens
(Sa\th{}^{\und*})a\o{}^{\und*})\\
&&=\sum a^{\und*}\t b^{\und*}\o\CR(b^{\und*}\t\tens
(Sa^{\und*}\o)a^{\und*}\th)=\sum a^{\und*}\t b^{\und*}\o\CR(b^{\und*}\t\tens
a^{\und*}\th)\CR(b^{\und*}\th\tens Sa^{\und*}\o)\\
&&=\sum \CR(b^{\und*}\o\tens a^{\und*}\t)b^{\und*}\t
a^{\und*}\th\CR(b^{\und*}\th\tens Sa^{\und*}\o)=\sum b^{\und*}\t
a^{\und*}\t\CR((Sb^{\und*}\o)b^{\und*}\th\tens Sa^{\und*}\o)\\
&&=b^{\und*}\und\cdot a^{\und*}}
using the properties (\ref{p8})(\ref{p9}) of $\CR$. It is evident that
$(\und*)^2=\id$ and condition (b) hold since $A$ is a Hopf $*$-algebra.
\endproof

Clearly, the proof of the preceding proposition holds for any real-dual
quasitriangular Hopf $*$-algebra $(A,\CR:A\tens A\to \C)$, with $\und A$ living
in the category of right $A$-comodules by the right adjoint coaction (this is
slightly more general than saying that $A$ is dual to some $H$).

\begin{propos} In the factorizable case of Proposition~2.6, the isomorphism
$Q:\und A\isom\und H$ due to the quantum Killing form is a $*$-isomorphism.
\end{propos}
\proof The map here is $a\in \und A$ maps to $Q(a)=(a\tens\id)(Q)$ where
$Q=\CR_{21}\CR_{12}$. In the real case we have $Q^{*\tens *}=Q$ and hence
\[Q(a^{\und *})=\sum <(Sa)^*,Q\uo>Q\ut=\sum
\overline{<a,Q\uo{}^*>}Q^\ut{}^*{}^*=(Q(a))^*.\]
This homomorphism $Q$ is a homomorphism of braided groups\cite{Ma:skl}, and we
see now that as such it is a $*$-homomorphism. In the (say finite-dimensional)
factorizable case it is, by definition an isomorphism.
\endproof

The process of transmutation of a dual quantum group into a braided one works
also at the level of matrix bialgebras. The braided version of the FRT
bialgebra $A(R)$\cite{FRT:lie} is the braided matrices $B(R)$ introduced in
\cite{Ma:exa}. These have matrix generators $u^i{}_j$ with relations and
braided-coproduct
\eqn{bmatrel}{R_{21}\vecu_1 R\vecu_2=\vecu_2 R_{21}\vecu_1 R,\quad \und\Delta
u=u\und\tens u,\ \und\eps u^i{}_j=\delta^i{}_j.}
The relations that arise here have been noted in various other contexts also.

\begin{corol} Let $R\in M_n\tens M_n$ be a matrix solution of the QYBE of {\em
real-type} in the sense
\[ \overline{R^i{}_j{}^k{}_l}=R^l{}_k{}^j{}_i,\quad {\rm i.e.,}\quad
\overline{R}^{t\tens t}=R_{21}.\]
Then
\[ u^i{}_j{}^{\und*}=u^j{}_i\]
makes $B(R)$ a real-braided matrix $*$-bialgebra in the sense of (a),(b),(d) of
Definition~2.4.
\end{corol}
\proof We first motivate the definitions. Thus, consider the dual quantum group
$A$ obtained as a quotient of the FRT bialgebra $A(R)$ with a matrix of
generators $t^i{}_j$ and relations $R\vect_1\vect_2=\vect_2\vect_1R$ as in
\cite{FRT:lie}. It is easy to see that if $R$ is of real-type then the compact
matrix-pseudogroup $*$-structure $t^i{}_j{}^*=St^j{}_i$ as in \cite{Wor:com} is
always compatible with the FRT relations -- we assume that it descends  to the
Hopf algebra level of $A$ also. Next, we have shown in \cite[Sec. 3]{Ma:qua}
(in some form) that $A(R)$ is indeed dual quasitriangular, i.e. there exists
$\CR:A(R)\tens A(R)\to\C$ obeying (\ref{p8})(\ref{p9}). It is $\CR(t^i{}_j\tens
t^k{}_k)=R^i{}_j{}^k{}_l$ extended according to (\ref{p8}), and we assume that
it descends to $\CR:A\tens A\to\C$. That this $\CR$ is real in the sense above
corresponds to
\[ \overline{R^i{}_j{}^k{}_l}=\overline{\CR(t^i{}_j\tens
t^k{}_l)}=\CR(t^k{}_l{}^*\tens t^i{}_j{}^*)=\CR(St^l{}_k\tens
St^j{}_i)=\CR(t^l{}_k\tens t^j{}_i)=R^l{}_k{}^j{}_i.\]
This is the reason for the condition stated. Assuming that $R$ is of real-type
we have for the corresponding $\und A$ as a quotient of $B(R)$ the
$*$-structure  $\und*=*\circ S=S^{-1}\circ *$ from Proposition~2.6. Here the
generators of $B(R)$ are identified with those of $A(R)$, $\vecu=\vect$ (but
not their products). Hence we have $u^i{}_j{}^{\und
*}=S^{-1}t^i{}_j{}^*=t^j{}_i=u^j{}_i$. This is the reason for the definitions.

More generally, we adopt this definition of $\und*$ at the level of the braided
matrices and verify directly that $B(R)$ obeys (a),(b) in Definition~2.4 as a
bialgebra in a braided category. Thus, to see that $\und*$ defines a
$*$-algebra structure on $B(R)$, we write out its defining relations;
\[ (R_{21}\vecu_1 R_{12}\vecu_2)^i{}_j{}^k{}_l\equiv R^k{}_a{}^i{}_b u^b{}_c
R^c{}_j{}^a{}_d u^d{}_l=u^k{}_a R^a{}_b{}^i{}_c u^c{}_d R^d{}_j{}^b{}_l\equiv
(\vecu_2 R_{21}\vecu_1 R_{12})^i{}_j{}^k{}_l.\]
Applying $\und*$ to this and using that $R$ is of real-type, we obtain the same
relations in the form
\[  (\vecu_2 R_{21}\vecu_1 R_{12})^j{}_i{}^l{}_k=u^l{}_d R^d{}_a{}^j{}_c
u^c{}_b R^b{}_i{}^a{}_k=R^l{}_b{}^j{}_d u^d{}_c
R^c{}_i{}^b{}_au^a{}_k=(R_{21}\vecu_1 R_{12}\vecu_2)^j{}_i{}^l{}_k.\]
This means that $\und*$ indeed extends to $B(R)$ as a $*$-algebra structure.
Finally, it obeys
\[ \und\Delta (u^i{}_j{}^{\und *})=\und\Delta u^j{}_i=u^j{}_k\tens
u^k{}_i=u^k{}_j{}^{\und*}\tens
u^i{}_k{}^{\und*}=\tau(\und*\tens\und*)\und\Delta u^i{}_j.\]
Finally, the condition (d) makes sense for action of a Hopf algebra $U(R)$ dual
to $A$ (and can be checked as such). Alternatively, it can more directly be
verified in a corresponding form for the right coaction of $A$, which takes the
form $\beta(u^i{}_j)=u^a{}_b\tens (St^i{}_a)t^b{}_j$. \endproof

If $A=G_q$ is the quantum group obtained from $A(R)$ for the quantum function
algebras associated to the standard semisimple Lie algebras as in
\cite{FRT:lie} (such as $SL_q(n)$) then the corresponding $\und A=BG_q$ is the
braided group of function algebra type. It is a quotient of the $B(R)$ modulo
suitable determinant-type relations. In this case there is a braided-antipode
and $\und*$ descends to a real-matrix braided group structure on $BG_q$. This
follows from Proposition~2.6 as explained in the preceding proof. For example,
the braided group $BSL_q(2)$ computed in \cite{Ma:exa} now becomes a real
$*$-braided group $BSU_q(2)$ with the $*$-structure above. Bearing in mind that
the braided groups of function algebra type are in a certain sense
braided-commutative, we see that we can think of $BSU_q(2)$ as the `ring of
functions' on some kind of braided 3-sphere. On the other hand, because the
transmuted $*$-structure is hermitian this $S^3$ has become a Lorentzian one
rather than the Euclidean one that we began with in the form of $SU_q(2)$. The
motivation for these interpretations is the well-known theorem of Gelfand and
Naimark that any commutative $C^*$-algebra is the functions on some locally
compact space. Finally, a corollary of Proposition~2.7 is

\begin{corol} Let $U_q(g)$ be a standard quantum group in FRT form with
generators $L=l^+Sl^-$. If the corresponding $R$-matrix is of real-type then
\[ L^i{}_j{}^*=L^j{}_i\]
defines a real-quasitriangular Hopf $*$-algebra structure on $U_q(g)$. This is
the case for the standard compact real form of $U_q(g)$ at real $q$.
\end{corol}
\proof From Proposition~2.7 we have $BU_q(g)\isom BG_q$ as $*$-braided groups.
The generator in the isomorphism corresponding to $u$ is $L=l^+Sl^-$ as
explained in \cite{Ma:skl}, giving the stated $*$-structure for $BU_q(g)$. But
this coincides with $U_q(g)$ as an algebra. A short computation shows that the
standard $R$-matrices as used in \cite{FRT:lie} are of real type if $q$ is real
and that this gives the standard compact $U_q(g)$ $*$-structure. It is also
known from general category-theoretical grounds that the standard compact
$*$-structures on $U_q(g)$ are real-quasitriangular\cite{Lyu:rea}. \endproof

Note that even for non-standard $R$ of real type, the definition
\eqn{lpm*}{l^\pm{}^i{}_j{}^*=Sl^{\mp}{}^j{}_i}
is compatible with the relations of the bialgebra $\tilde U(R)$ in \cite[Sec.
3]{Ma:qua} and if it descends to the Hopf algebra quotient $U(R)$, will define
a Hopf $*$-algebra structure with $L^*=L^{t}$ as above. Here the $*$-structure
on $U_q(g)$ or $U(R)$ forms a Hopf $*$-algebra. It specifies a real form of the
quantum group as a symmetry (such as angular momentum).

\section{$D(H)$ as a Quantum Algebra of Observables}

In this section we apply the algebraic results of the preceding section to give
a quantum-mechanical picture of the quantum double. We begin by recalling the
relevant definition of quantization according to the cross product construction
of Mackey. In the classical case it can be understood in more conventional
terms of quantization of Poisson brackets etc, but this need not concern us
now. It has an obvious generalization to deal with $q$-groups on $q$-spaces
which we will then use.

Thus, let $X$ be a manifold and $G$ a Lie group with Lie algebra $g$, acting on
$X$. We consider the quantization of a particle on $X$ moving (on orbits)
according to the action of $G$ (in nice cases there is a metric on $X$ such
that this motion is geodesic motion). The position observables are functions
$C(X)$ (for example, the $C^*$-algebra of functions vanishing at infinity when
$X$ is locally compact), while the momentum observables are elements $\xi$ of
$g$. They are constants of the motion. Quantization of the system means to find
a $*$-algebra $\CA$ (usually a $*$-subalgebra of bounded operators on a Hilbert
space) containing the position and momentum observables as $*$-subalgebras such
that
\eqn{heis}{\widehat{e^{\tau\xi}}\widehat
f\widehat{e^{-\tau\xi}}=\widehat{\alpha_{e^{\tau\xi}}(f)},\quad
\alpha_{e^{\tau\xi}}(f)(x)=f(\alpha_{e^{-\tau\xi}}(x))}
where $\alpha$ denotes the action on $X$ and also the induced action on $f\in
C(X)$, and $\hat{\ }$ denotes the embeddings of the position and momentum
observables in $\CA$ (the quantization maps).
This requirement is just a co-ordinate-free version of the usual Heisenberg
commutation relations, for it says that conjugation by the exponentiated
momentum in the quantum algebra implements translation on the underlying
position space.

On the other hand, there is a universal solution to this quantization problem.
It is the cross product algebra
$C(X)\cocross G$ by $\alpha$. Here, for convenience, we have shifted from the
Lie algebra $g$ (or its enveloping algebra $U(g)$) to the group $C^*$-algebra
$\C G$  generated (roughly speaking) by elements of $G$. This is for
convenience since such exponentiated elements tend to be represented by bounded
operators rather than unbounded ones. This is a standard construction in the
theory of operator algebras (there is also a von Neumann algebra version) and
$X$ and $G$ need only be locally compact rather than smooth manifolds. Any
other quantization $\CA$ is merely a $*$-representation of this cross product.
The construction is roughly equivalent to Mackey's theory of systems of
imprimitivity\cite{Mac:ind}.

Many authors since Mackey have independently rediscovered this as a natural
method of quantization of homogeneous spaces. See \cite{Ma:phy} for the version
relevant to the present considerations of Hopf algebras. Specifically, we noted
there that the notion of cross product is well-known to make sense when $G$ is
replaced by a non-cocommutative quantum group $H$ such as (the operator-algebra
version of) $U_q(g)$ and $C(X)$ by some non-commutative $*$-algebra. We have
already given a canonical example of this type\cite[Example 4.7]{Ma:phy} where
the cross product quantization happens again to be a Hopf algebra. We want to
pursue this now for the quantum double.

Thus, we take for our `position functions' some non-commutative $*$-algebra
$B$, typically a $q$-deformation of some $C(X)$. For simplicity, we assume it
is unital. We suppose that there is an action $\alpha=\la$ of a Hopf
$*$-algebra $H$ on $B$ such that
\eqn{leftuni}{h\la (bc)=\sum (h\o\la b)(h\t\la c),\quad h\la 1=\eps(h)1,\qquad
(h\la b)^*=(Sh)^*\la b^*,\quad h\in H,\ b\in B.}
The first conditions are the Hopf algebra analogue of the condition that $H$
acts `by automorphisms' of $B$ and means that we can form a cross product
algebra $B\cocross H$ as explained in the preliminaries. It indeed has $B,H$ as
sub-algebras and cross relations in the equivalent form
\eqn{crosscom}{\sum (1\tens h\o)(b\tens 1)(1\tens S h\t)=(h\la b\tens 1),\quad
h\in H,\ b\in B.}
The last condition in (\ref{leftuni}) is a `unitarity' condition arising from
the requirement that $B\cocross H$ is a $*$-algebra with $B,H$ as
$*$-subalgebras (just apply $*$ to (\ref{crosscom}) and require that we obtain
(\ref{crosscom}) again for $(Sh)^*,b^*$). This condition has already been noted
in \cite{Egu:mec} where $B$ is the quantum sphere and $H=U_q(su_2)$. When it
holds, $B\cocross H$ becomes a $*$-algebra. We have also encountered both
conditions in Section~2. The quantization maps $\widehat b=b\tens 1$ and $\hat
h=1\tens h$ are then embeddings of $B,H$ and in terms of them, (\ref{crosscom})
becomes
\eqn{qheis}{ \sum \widehat{h\o}\widehat b\widehat{Sh\t}=\widehat{h\la b},\qquad
b\in B,\ h\in H}
as a generalization of (\ref{heis}).

We have already seen in Proposition~2.5 that the quantum adjoint action of $H$
on itself obeys (\ref{leftuni}), so $H\cocross H$ by this is a $*$-algebra.
Likewise for several other canonical actions such as the left coregular action
of $H^*$ on $H$. The latter has quantization given by the cross product
$H^*\cocross H$ (generalizing the Weyl algebra on a group). Here $H^*\cocross
H\isom \Lin(H)$ but is not in general a Hopf algebra. Moreover, we have seen in
Proposition~2.6 that the coadjoint action of $H$ on $\und A$ also precisely
obeys the conditions (\ref{leftuni}). This is the form that we need.

\begin{corol}
If $H$ is a real-quasitriangular Hopf $*$-algebra in the sense $\CR^{*\tens
*}=\CR_{21}$ then $\und A\cocross H$ is a $*$-algebra cross product as above (a
quantization). It is a Hopf algebra isomorphic to $D(H)$.
\end{corol}
\proof  This follows at once from Proposition~2.6. Here $\und A$ has its
$*$-algebra structure $\und *$ and $H$ acts by the quantum coadjoint action.
The $*$-structure on $\und A\cocross H$, which is the one relevant to the
generalized Mackey quantization makes $\und A$ and $H$ $*$-subalgebras.
The isomorphism $\theta:\und A\cocross H\to D(H)$ was obtained in
\cite{Ma:dou}\cite{Ma:skl} as
\eqn{theta}{\theta(a\tens h)=\sum a\o <\CR\uo,a\t>\tens \CR\ut h,\qquad
\theta^{-1}(a\tens h)=\sum a\o<S\CR\uo,a\t>\tens\CR\ut h.}
This isomorphism induces a $*$-algebra structure in $D(H)$ but note that it is
quite different from that in Example~2.1 where the subalgebras $H^{*\rm op}$
and $H$ were $*$-subalgebras. In fact, the quantum mechanical $*$-structure
from $\und A\cocross H$ does not naturally form a Hopf $*$-algebra -- this is
evident from the explicit form of the coproduct of $\und A\cocross H$ given in
\cite{Ma:skl} as
\eqn{bosdelta}{ \Delta (a\tens h)=\sum a\o\tens \CR\ut h\o\tens \CR\uo\la
a\t\tens h\t.}
This is a cross coproduct by coaction $\beta(a)=\CR_{21}\la a$ where the second
factor of $\CR_{21}$ acts on $a$ in $\und A$ by the quantum coadjoint action.
\endproof

We conclude from this that $D(H)$ in this case is the algebra of observables
for the quantization of a particle on the non-commutative space $B=\und A$ with
momentum given by the quantum group $H$. This is a very general result. We
concentrate now on the matrix case where $H=U_q(g)$ is in FRT form with
generators $l^\pm$ and dual $G_q$ of  function algebra type. We have already
recalled the braided groups $BG_q$ from \cite{Ma:exa} as quotients of the
braided-matrices $B(R)$ with matrix generator $\vecu$ (see the end of
Section~2).

\begin{corol} $BG_q\cocross U_q(g)$ with $q$ real is a $*$-algebra cross
product describing a quantum particle with positions observables $BG_q$ and
momentum $U_q(g)$, and is a Hopf algebra isomorphic to $D(U_q(g))$.
Explicitly, it has matrix generators $u$ of $BG_q$ and $\l^\pm$ of $U_q(g)$
with cross relations and coproduct,
\[ Rl_2^+u_1=u_1 Rl_2^+,\quad R_{21}^{-1}l_2^-u_1=u_1 R_{21}^{-1} l_2^-,\quad
\Delta l^\pm=l^\pm\tens l^\pm,\quad \Delta u=\sum u\CR\ut\tens \CR\uo\la u\]
where $\CR=\sum\CR\uo\tens\CR\ut$ is the universal $R$-matrix and $\la$ is the
quantum adjoint action. The $*$-structures are $u^i{}_j{}^{\und *}=u^j{}_i$,
$l^\pm{}^i{}_j{}^*=S l^\mp{}^j{}_i$.
\end{corol}
\proof This is a special case of the last corollary. The quantum coadjoint
action on $BG_q$ coincides on the matrix generators with that on $G_q$ (and
with the quantum adjoint action on $L$ in $U_q(g)$). It comes out as in
\cite{Ma:skl} as
\eqn{mataction}{ l^+{}^k{}_l\la u^i{}_j=R^{-1}{}^i{}_a{}^k{}_b u^a{}_c
R^c{}_j{}^b{}_l,\quad l^-{}^i{}_j\la u^k{}_l=R^i{}_a{}^k{}_b u^b{}_c
R^{-1}{}^a{}_j{}^c{}_l}
giving the relations shown, while the isomorphism (\ref{theta}) of the
resulting cross product with $D(U_q(g))$ comes out as
\eqn{mattheta}{\theta(1\tens l^\pm)=1\tens l^\pm,\quad \theta(u^i{}_j\tens
1)=t^i{}_a\tens Sl^-{}^a{}_j.}
These steps are similar to those in \cite[Sec. 4]{Ma:skl}. Also, from
Corollary~3.1, it follows that we have a $*$-algebra cross product when $q$ is
real. We can also verify this directly using the reality  property of the
$R$-matrix in Corollary~2.8,
\[(l^+{}^k{}_l\la u^i{}_j)^{\und
*}=\overline{R^{-1}{}^i{}_a{}^k{}_b}(u^a{}_c){}^{\und *}\overline{
R^c{}_j{}^b{}_l}=R^l{}_b{}^j{}_c u^c{}_a R^{-1}{}^b{}_k{}^a{}_i=l^-{}^l{}_k\la
u^j{}_i=(Sl^+{}^k{}_l)^*\la (u^i{}_j)^{\und *}\]
and similarly for $l^-$. Note that using the axioms for $\CR$, one can also
compute the coproduct further as
\eqn{matcoprod}{ \und\Delta u=(\sum u\CR\ut\tens \CR\uo u)\CR^{-1}_{21}}
where the $\CR$ live in the tensor square of $U_q(g)$.
\endproof

This completes the general theory. To all the compact Lie group deformations we
 have associated quantum systems isomorphic as algebras to the quantum double.
The position space is not an ordinary group or quantum group but a braided one.
Moreover, its $*$-structure needed for our interpretation is transformed in the
process from unitary to hermitian in its general character. We conclude now
with the full details in the simplest case, namely $g=su_2$.

Firstly, the position observables is given by the algebra $BSU_q(2)$ with four
generators (the matrix `coordinates') and relations\cite{Ma:exa}
\eqn{bm2a}{ba=q^2ab,\qquad ca=q^{-2}ac,\qquad da=ad,\qquad
bc=cb+(1-q^{-2})a(d-a)}
\eqn{bm2b}{db=bd+(1-q^{-2})ab,\qquad cd=dc+(1-q^{-2})ca}
\eqn{bm2-det}{ad-q^2 cb=1}
and equipped now with hermitian $*$-structure as quoted in the Introduction. If
we write new self-adjoint generators
\eqn{mink}{x_0=qd+q^{-1}a,\quad x_1={b+c\over 2},\quad x_2={b-c\over 2i},\quad
x_3=d-a}
then $x_0$ (the time direction) is central and the left hand side of
(\ref{bm2-det}) is
\eqn{minkmetric}{ ad-q^2cb={q^2\over(q^2+1)^2}x_0^2-q^2x_1^2-q^2 x_2^2-
{(q^4+1)q^2\over 2(q^2+1)^2}x_3^2+\left({q^2-1\over q^2+1}\right)^2{q\over 2}
x_0x_3}
so that setting this braided-determinant equal to 1 means that our position
observables are a q-deformation of a hyperboloid or 3-sphere in
Minkowski-space. This $*$-algebra can be denoted $BS^3_q$ for this reason.

Secondly, our momentum group is $U_q(su_2)$ with its usual $*$-structure. Its
action on the position observables comes out from the above theory
as\cite{Ma:exa}
\eqn{su2actiona}{ X_+\la\pmatrix{a&b\cr
c&d}=\pmatrix{-q^{3\over2}c&-q^{\h}(d-a)\cr 0 & q^{-\h}c}\to [\pmatrix{a&b\cr
c&d},\pmatrix{0&1\cr 0&0}]}
\eqn{su2actionb}{ X_-\la\pmatrix{a&b\cr c&d}=\pmatrix{q^{\h}b&0\cr
q^{-\h}(d-a)& -q^{-{3\over 2}}b}\to [\pmatrix{a&b\cr c&d},\pmatrix{0&0\cr
1&0}]}
\eqn{su2actionc}{H\la\pmatrix{a&b\cr c&d}=\pmatrix{0&-2b\cr -2c& 0}\to
[\pmatrix{a&b\cr c&d},\pmatrix{1&0\cr 0&-1}]}
where the limits are as $q\to 1$. These make clear that our action is a
q-deformation of the one induced on the generators of the ring of co-ordinate
functions by the adjoint action.
At the group level this is the usual action induced on $C(S^3_{\rm Lor})$ by
conjugation in $SL_2$. The orbits of this are spatial spheres.

The quantization of this system as a $*$-cross product is then isomorphic to
the quantum double $D(U_q(su_2))$. Thus the quantum double should be viewed as
a $q$-deformed version of quantum motion on spheres. Of course, in the
$q$-deformed setting there are neither actual points nor actual orbits in the
usual sense. Moreover, one has to quantize the spheres  together as a foliation
of a hyperboloid in Minkowski space (rather than the more obvious setting of
spheres in Euclidean space) for this interpretation to work in the q-deformed
case.

This complete the details of the simplest model as discussed in
(\ref{double.int}) in the introduction. We note that the $q$ that enters into
the non-commutation relations (\ref{bm2a})-(\ref{bm2b}) should be thought of as
due to braid statistics and not due to any quantization\cite{Ma:exa}. This is
because the braided-group function algebra here is in a certain sense
braided-commutative (cf the super-commutativity of the co-ordinate ring for
supergroups) and this is expressed by these relations. This point of view
justifies our use of $BSU_q(2)=BS_q^3$ as the {\em classical} position space
before quantization. The quantum group $U_q(su_2)$ is likewise to be viewed as
some kind of deformation, rather than quantization, of the classical angular
momentum $U(su_2)$. Although $BSU_q(2)$ and $U_q(su_2)$ are the same as
$*$-algebras, their interpretation and coproducts are quite different. The
quantum double is then the Mackey-type quantization of this $q$-deformed
system.

Finally, we note that the models above can be easily extended using the same
techniques. For example, we can relax (\ref{bm2-det}) and thereby take for our
position space a q-deformation of Minkowski space. In this case the algebra is
isomorphic to the degenerate Sklyanin algebra as explained in \cite{Ma:skl}
where also the action of $U_q(su_2)$ is given.
The momentum group can also be extended from $U_q(su_2)$ to its quantum group
complexification. For this one can take the quantum double $D(U_q(su_2))$
regarded this time as a q-deformation of the Lorentz group\cite{PodWor:def}.
This system will be explored elsewhere.

\section{The dual of $D(H)$ as another quantum algebra of observables}

The striking consequence of an algebra of observables being a Hopf algebra is
that its dual (defined suitably in the infinite-dimensional case) is also a
Hopf algebra. But is this the algebra of observables of some other quantum
system, dual to the first? This was the case in the models in
\cite{Ma:phy}\cite{Ma:mat}\cite{Ma:hop}\cite{Ma:pla} and we will see that it is
the case here also for the quantum double. The main reason is that $\und
A\cocross H$ (isomorphic to the quantum double) has a crossed structure not
only as an algebra but also as a coalgebra.

The axioms for the cross coproduct coalgebra structure were recalled in the
Preliminaries and are based on a left comodule coalgebra structure obeying
(\ref{p4}) and dual to those for a left cross product algebra $B\cocross H$ as
in (\ref{p1}). Here it is an elementary fact that a left $H$-module algebra $B$
with action $\la:H\tens B\to B$ dualises (at least in the finite-dimensional
case) to a left $A$-comodule coalgebra $C$ with coaction $\beta:C\to A\tens C$
where $C$ is the coalgebra dual to the algebra $B$ and $A$ a Hopf algebra dual
to $H$. Likewise vice versa. Hence if, as for the double, $B$ is both a left
$H$-module algebra and a left $H$-comodule coalgebra then $C$ is both a left
$A$-comodule coalgebra and left $A$-module algebra. Moreover, if $B\cocross H$
with the resulting crossed algebra and coalgebra structures turns out to be a
Hopf algebra\cite{Rad:str}, clearly the dual algebra and coalgebra $C\cocross
A$ is then also a Hopf algebra. Applying this to the full Hopf algebra
structure underlying Corollary~3.1 we obtain,

\begin{propos}  If $A$ is an antireal dual-quasitriangular Hopf algebra dual to
$H$ then $(\und H)^{\rm op/op}\cocross A$ is a $*$-algebra cross product,
isomorphic to $D(H)^*$ as a Hopf algebra. Here $(\und H)^{\rm op/op}$ denotes
the braided group $\und H$ with the opposite product and coproduct.
\end{propos}

The detailed proof here requires us to compute the relevant dual as $(\und
A)^*=(\und H)^{\rm op/op}$, as well as to analyse the condition needed for the
resulting action of $A$ (as obtained by dualization) to obey the unitarity
condition as in (\ref{leftuni}) in order to have a $*$-algebra cross product.
We do this now, but in some cleaner conventions that avoid the use of $(\
)^{\rm op/op}$ and corresponding complexities in the formulae: these are all
avoided if in the course of dualization we make a further switch from
left-modules and comodules to right-modules and comodules. Thus, for our
explicit presentation of results in this section, we now switch over to these
right-handed conventions.

Firstly, $C$ is a right $A$-module algebra (where $A$ is a Hopf algebra) if
$\ra$ is a right action and
\eqn{rightmodalg}{(cd)\ra a=\sum (c\ra a\o)(d\ra a\t),\quad 1\ra a=\eps(a)1.}
In this case, there is a cross product algebra $C\cocross A$ with
\eqn{rightcross}{ (c\tens a)(d\tens a')=\sum (c\ra a'\t)d\tens a(a'\o).}
Next, C is a right $A$-module coalgebra if it is a coalgebra and $\ra:C\tens
A\to C$ is a coalgebra map. It is a right $A$-comodule algebra if it is an
algebra and there is a right comodule $\beta:C\to C\tens A$ which is an algebra
map.
Finally, it is a right $A$-comodule coalgebra if it is a coalgebra and $\beta$
is a right comodule obeying
\eqn{rightcomodcoalg}{(\Delta_C\tens\id)\beta(c)=\sum c\o\bo\tens c\t\bo\tens
c\o\bt c\t\bt,\qquad (\eps_C\tens\id)\beta(c)=1\eps_C(c)}
where $\beta(c)=\sum c\bo\tens c\bt\in C\tens A$.
In the last case there is a cross coproduct coalgebra $C\cocross A$ with
\eqn{rightcocross}{ \Delta (c\tens a)=\sum c\o\bo\tens a\o\tens c\t\tens a\t
c\o\bt.}

This summarises the right-handed version of the formulae in the Preliminaries.
Also, if $A$ is a Hopf $*$-algebra and $C$ an $A$-module algebra and
$*$-algebra, for the cross product to be a $*$-algebra containing $A,C$ as
$*$-subalgebras, we require
\eqn{rightuni}{ (c\ra a)^*=c^*\ra (Sa)^*.}
This is the right-handed analogue of the unitarity condition in
(\ref{leftuni}).

\begin{propos} cf\cite{Rad:str} Let $A$ be a Hopf algebra. If $(C,\und\Delta)$
is both a right $A$-module algebra and coalgebra, and a right $A$-comodule
algebra and coalgebra and
\[ \und \Delta (cd)=\sum c\Bo d\Bo\bo\tens (c\Bt\ra d\Bo\bt)d\Bt,\quad \sum
(c\ra a\t)\bo\tens a\o(c\ra a\t)\bt=\sum c\bo\ra a\o\tens c\bt a\t\]
along with $\und\Delta 1=1\tens 1$ etc, and a convolution-inverse of the
identity $C\to C$, then $C\cocross A$ by the cross product and coproduct, is a
Hopf algebra.
\end{propos}
\proof This is nothing more than a right-handed version of an observation of
Radford in \cite{Rad:str}. Namely, as $C$ is an $A$-module algebra, we have an
associative cross product $C\cocross A$, and since it is an $A$-comodule
coalgebra, we have a coassociative cross coproduct. The remaining conditions
ensure that these fit together to form a Hopf algebra. This is an elementary
computation. In fact, there is a simple way to do all these dualization
computations easily, by diagrammatic methods\cite{Ma:introp}. \endproof

To understand why these right-handed conventions are appropriate in the dual,
note first that the assertion that $B$ is a left $H$-module algebra or
coalgebra is equivalent to the assertion that its product or coproduct maps are
intertwiners for the action of $H$, that is that $B$ lives in the category of
$H$-modules. In this setting  $B$ has a natural dual $B^\star$ which also lives
in the same category. As a linear space it coincides with the usual dual $B^*$,
but has the opposite coproduct or product to the usual one,
\eqn{bstar}{B^\star=B^*{}^{\rm op/op}.}
Moreover, $B^\star$ is again a left $H$-module under the action $<h\la
f,b>=<f,(Sh)\la b>$ for all $b\in B$, $f\in B^\star$.

\begin{lemma} Let $H$ be a finite-dimensional Hopf algebra with dual $A$. If
$B$ is a finite-dimensional left $H$-module algebra and $H$-comodule coalgebra,
such that $B\cocross H$ is a Hopf algebra, then $C=B^\star$ is in the situation
of Proposition~4.2 and
\[ C\cocross A=(B\cocross H)^\star\isom (B\cocross H)^*.\]
\end{lemma}
\proof  One can see easily that if $B$ is a left $H$-module (algebra,
coalgebra) then $B^\star$ is a left $H$-module (coalgebra, algebra). Also, a
left $H$-comodule (algebra, coalgebra) defines a right $A$-module (algebra,
coalgebra) by usual dualization, where $A$ is dual to $H$. Likewise, a left
$H$-module (algebra, coalgebra) is equivalently a right $A$-comodule (algebra,
coalgebra). Combining these elementary observations, we see that a left
$H$-comodule (algebra, coalgebra) $B$ has as its categorical dual $B^\star$ a
right $A$-module (coalgebra, algebra), and a left $H$-module (algebra,
coalgebra) has as its categorical dual a right $A$-comodule (coalgebra,
algebra). Explicitly, the two are related by
\eqn{dualaction}{ \sum <b,c\bo><h,c\bt>=<c,(Sh)\la b>,\quad <b,c\ra(Sa)>=\sum
<a,b\bo><c,b\bt>}
for all $b\in B$, $c\in B^\star$, $h\in H$ and $a\in A=H^*$. From this and an
elementary computation, one sees that if $B\cocross H$ is a Hopf algebra (its
module and comodules structures obey the original left-handed
form\cite{Rad:str} of the conditions in Proposition~4.2) then the right handed
$C\cocross A$ is also a Hopf algebra by the proposition. Its natural pairing is
with $(B\cocross H)^\star$ by the map
\[ C\cocross A\to (B\cocross H)^\star,\quad c\tens a\mapsto <c\tens S^{-1}a,\
>;\]
\align{&&\nqquad< (b\tens h)(b'\tens h'),c\tens S^{-1}a>=\sum <b,c\Bo><h\o\la
b',c\Bt><h',S^{-1}a\o><h\t,S^{-1}a\t>\\
&&=\sum <b',c\Bo\bo><S^{-1}h\o,c\Bo\bt><b,c\Bt><h',S^{-1}a\o><h\t,S^{-1}a\t>\\
&&=\sum <b'\tens h'\tens b\tens h, c\Bo\bo\tens S^{-1}a\o\tens c\Bt\tens
S^{-1}(a\t c\Bo\bt)>\\
&&=<b'\tens h'\tens b\tens h,(\id\tens S^{-1}\tens\id\tens
S^{-1})\Delta_{C\cocross A} c\tens a>\\
&&\nqquad <b\tens h,(\id\tens S^{-1})((c\tens a)(c'\tens a'))>=\sum <b\tens
h,(c\ra a'\t)c'\tens (S^{-1}a'\o)S^{-1}a>\\
&&=\sum <b\Bo,c'><b\Bt,c\ra a'\t><h,(S^{-1}a'\o)S^{-1}a>\\
&&=\sum <b\Bo,c'><S^{-1}a'\t,b\Bt\bo><b\Bt\bt,c><h\o,S^{-1}a'\o><h\t,
S^{-1}a>\\
&&=\sum <c'\tens S^{-1}a'\tens c\tens S^{-1}a,b\Bo\tens b\Bt\bo h\o\tens
b\Bt\bt\tens h\t>\\
&&=<c'\tens S^{-1}a'\tens c\tens S^{-1}a,\Delta_{B\cocross H} b\tens h>}
where $\Delta_B b=\sum b\Bo\tens b\Bt$ and $\Delta_C c=\sum c\Bo\tens c\Bt$
with $C=B^\star$ as in (\ref{bstar}). Note that because $B\cocross H$ is a Hopf
algebra, its usual dual $(B\cocross H)^*$ and $(B\cocross H)^{*\rm op/op}$ are
isomorphic via the antipode of $(B\cocross H)^*$.
\endproof

There is a similar result in the setting of dually-paired Hopf algebras. This
takes care of the general constructions of which the quantum double and its
dual (as we will see) are  examples. We are now ready to study the situation
when $H$ is quasitriangular, with quasitriangular structure $\CR$, or slightly
more generally  when $A$ is dual quasitriangular with $\CR\in (A\tens A)^*$.
The way that the cross product and coproduct  structure of $D(H)$ was found in
\cite{Ma:dou} was to show that if $B$ is an $H$-module (algebra, coalgebra)
then $\CR$ can be used to also make $B$ an $H$-comodule (algebra, coalgebra)
and this led to $B\cocross H$ as a Hopf algebra. The corresponding lemma in the
dual language of $(A,\CR)$ is

\begin{lemma} cf\cite{Ma:dou}\cite{Ma:bos} Let $(A,\CR)$ be dual
quasitriangular. If $C$ is a right $A$-comodule (algebra, coalgebra) then
\[ c\ra a=\sum c\bo\CR(c\bt\tens a).\]
makes $C$ also a right $A$-module (algebra, coalgebra) and the second condition
stated in Proposition~4.2 is satisfied. If $C$ is a Hopf algebra living in the
braided category of $A$-comodules then $C\cocross A$ is a Hopf algebra by
Proposition~4.2.
\end{lemma}
\proof The first part is nothing other than a dual version of an observation
first made in \cite[Prop. 3.1]{Ma:dou}, and in any case follows at once from
(\ref{p8})-(\ref{p9}). The second part is nothing other than a dual version of
\cite[Thm 4.1]{Ma:bos} (and corresponds to turning the diagram-proofs there
upside down). It is also easy to see directly: For $C$ a Hopf algebra living in
the braided category of right $A$-comodules the condition that $\und\Delta:C\to
C\und\tens C$ is an algebra homomorphism to the braided tensor
product\cite{Ma:bg},
\eqn{bhopfdual}{\und\Delta(cd)=\sum c\Bo d\Bo\bo\tens c\Bt\bo d\Bt
\CR(c\Bt\bt\tens d\Bo\bt)}
reduces in the present context to the first condition of Proposition~4.2.
Hence, we have a biproduct $C\cocross A$ forming a Hopf algebra. Note that the
usual dualization of \cite[Thm 4.1]{Ma:bos} is obtained trivially by dualizing
the relevant structure maps (and adapting to the infinite-dimensional case),
and does not require the right-handed theory above. \endproof

\begin{lemma} cf\cite{LyuMa:bra} Let $H$ be a quasitriangular Hopf algebra with
dual $A$ and associated braided groups $\und H,\und A$. Then we have as Hopf
algebras in the braided category of $H$-modules,
\[  \und H\isom (\und A)^\star.\]
\end{lemma}
\proof We consider the map $\und H\to (\und A)^\star$ by $b\mapsto <Sb,\ >$ and
use (\ref{p13})-(\ref{p14}) to do all our computations in terms of the usual
Hopf algebras $H,A$. Then
\align{<Sb,a\und\cdot a'>\nqquad &&=\sum
<Sb\t,a\t><Sb\o,a'\t>\CR((Sa\o)a\th\tens Sa'\o)\\
&& =\sum <(S\CR\uo\o)(Sb\t)\CR\uo\t,a><(S\CR\ut)Sb\o,a'>\\
&&=\sum <S(b\o\CR\ut),a'><S(S^{-1}\CR\uo\t b\t \CR\uo\o),a>\\
&&=\sum <S(b\o S\CR\ut),a'><S(\CR\uo\o b\t S\CR\uo\t),a>=<(S\tens S)\und\Delta
b,a'\tens a>}
as required. The pairing $<Sb\tens Sb',\und\Delta a>=<S(b'b),a>$ is easier
since the product of $\und H$ and the coproduct of $\und A$ coincide with the
usual ones. Likewise for the pairing of the units and counits.
We also check that the map $<S(\ ),\ >$ is indeed a morphism in the category of
$H$-modules,
\align{(<S(h\la b),\ >)(a)&&=<S(h\la b),a>=\sum <h\o b Sh\t, Sa>\\
=&&\sum <Sh\o,a\th><b,Sa\t><S^2h\t,a\o>\\
=&&\sum <Sb,a\bo><Sh,a\bt>=<Sb,(Sh)\la a>=(h\la<Sb,\ >)(a)}
where the natural action on $f\in (\und A)^\star$ is $(h\la f)(a)=f((Sh)\la a)$
as explained above. \endproof

Putting together these various preliminary facts, we obtain

\begin{propos} Let $H$ be an antireal-quasitriangular Hopf algebra with dual
$A$ as in Section~2, and $\und H$ the braided group associated to $H$ (here
$\und H$ coincides with $H$ as a $*$-algebra). Then $\und H\cocross A$ is a
(right handed) $*$-algebra cross product, and is a Hopf algebra isomorphic to
$D(H)^*$. Here the action of $A$ on $\und H$ is
\[ b\ra a=\sum \CR\uo\la b <\CR\ut,a>\]
where $\la$ is the quantum adjoint action.
\end{propos}
\proof The general theory has been established above in the required form. We
conclude from this that $\und H\cocross A$ is  a right-handed cross product and
coproduct, and forms a Hopf algebra dual to the left-handed $\und A\cocross H$
in Corollary~3.1. It remains only to see how the action, coaction and
isomorphism look explicitly and to see the condition needed for a $*$-algebra
structure for the cross product. From (\ref{dualaction}) we have
\align{&&\nqquad <S(b\ra a),f>=<<Sb,\ >\ra a,f>=\sum <S^{-1}a,\CR\ut><\CR\uo\la
f,Sb>\\
&&=\sum <S^{-1}a,\CR\ut><f\t,Sb><\CR\uo, (Sf\o)f\th>\\
&&=\sum <\CR\ut,a><(Sf)\t,b><\CR\uo,(Sf)\o S(Sf)\th>\\
&&=\sum <\CR\ut,a><S(\CR\uo\o b S\CR\uo\t),f>}
for all $f\in \und A$. From this we conclude the form shown. In a similar way
we can obtain the right coaction of $A$ on $\und H$ dual via (\ref{dualaction})
to the left coadjoint action of $H$ on $\und A$. It comes out as determined by
\eqn{rightco}{ \sum b\bo\tens <b\bt,h>=h\la b\qquad \forall h\in H,\ b\in \und
H}
and is necessarily related to the right action $\ra$ as in Lemma~4.4 with
$C=\und H$.

In a similar way one can trace through the definitions to deduce from the dual
of (\ref{theta}) the isomorphism ${\hat\theta}:\und H\cocross A\to D(H)^*$ as,
\eqn{cotheta}{{\hat\theta}(b\tens a)=<\CR\ut e_a\o,a>(S\CR\uo) (e_a\t\la
b)\tens f^a}
where $\{e_a\}$ is a basis of $H$ and $\{ f^a\}$ is a dual one. This is
obtained from the isomorphisms $\und H\cocross A \isom (\und A)^\star\cocross A
\isom  D(H)^\star \isom  D(H)^*$ by Lemma~4.5, Lemma~4.3 and Corollary~3.1. The
first isomorphism is via $<S(\ ),\ >$, the second via $<\id\tens S^{-1},\ >$,
the third via the adjoint of (\ref{theta}) and the last via the adjoint of the
antipode of $D(H)$. Thus,
\align{&&\nqquad<{\hat\theta}(b\tens a),a'\tens h>=<b\tens a,(S\tens
S^{-1})\circ\theta^{-1}\circ S_{D(H)}(a'\tens h)>\\
&&=<b\tens a,(S\tens S^{-1})\circ\theta^{-1}((1\tens Sh)(S^{-1}a'\tens 1))>\\
&&=\sum <b\tens a,(S\tens S^{-1})\circ\theta^{-1}(S^{-1}a'\t\tens
Sh\t)><Sh\th,a'\th><h\o,a'\o>\\
&&=\sum <b\tens a,a'\th\tens h\t\CR\ut><h\o,
a'\o><S\CR\uo,a'\t><Sh\th,a'_{(4)}>\\
&&=\sum <h\o (S\CR\uo)bS h\th,a'><a,h\t\CR\ut>\\
&&=\sum <(S\CR\uo)h\t bS h\th,a'><a,\CR\ut h\o>=\sum <a,\CR\ut h\o>
<(S\CR\uo)(h\t\la b),a'>}
from which we deduce (\ref{cotheta}) at least in the finite-dimensional case.

Finally, we suppose that $\CR$ is antireal. Then
\align{ (b\ra a)^*\nqquad && =\sum \overline{<\CR\ut,a>}(\CR\uo\la b)^*=\sum
<(S\CR\ut)^*,a^*>(S\CR\uo)^*\la b^*\\
&&=\sum <\CR\ut{}^*,a^*>\CR\uo{}^*\la b^*=\sum <\CR\ut,a^*>(S\CR\uo)\la b^*\\
&&=\sum <\CR\ut,S^{-1}a^*>\CR\uo\la b^*=b^*\ra S^{-1}(a^*)=b^*\ra (Sa)^*}
as required. Hence, if $A$ is antireal dual-quasitriangular, then $\und
H\cocross A$ is a $*$-algebra cross product. \endproof

Note that the condition on $\CR$ for a $*$-algebra cross product here is
different from the one in Corollary~3.1.

\begin{example} $BU_q(sl(2,\R))\cocross SL_q(2,\R)$ with $|q|=1$ is a
$*$-algebra cross product describing a quantum particle with momentum
$SL_q(2,\R)$ and position observables $BU_q(sl(2,\R))$, and is a Hopf algebra
isomorphic to the dual of $D(U_q(sl_2))$. Explicitly, it has matrix generators
$L$ of $BU_q(sl(2,\R))$ and $t$ of $SL_q(2,\R)$ with cross relations and
coproduct
\[ L_1 t_2=t_2 R^{-1}L_1 R,\qquad \Delta t=t\tens t,\qquad \Delta
L^i{}_j=L^a{}_b\tens (St^i{}_a)t^b{}_c L^c{}_j.\]
The action and coaction here are $L_1\ra t_2=R^{-1} L_1 R$ and $\beta(L)=t^{-1}
L t$ in the usual notations.
The same result holds for any other $U_q(g)$ for which there is an antireal
$*$-structure.
\end{example}
\proof This is a special case of the above Proposition~4.6.  We compute the
action there as
\eqn{rightmatact}{L^i{}_j\ra t^k{}_l=\CR\uo\la
L^i{}_j<\CR\ut,t^k{}_l>=l^+{}^k{}_l\la L^i{}_j}
as already given in the proof of Corollary~3.2. The coaction follows from
(\ref{rightco}) as
\eqn{rightmatcoact}{ \beta(L^i{}_j)=L^a{}_b\tens (St^i{}_a)t^b{}_j}
since this dualises to the left quantum adjoint action of $H$ on $\und H$. This
is usually written compactly as conjugation by $\vect$. From (\ref{rightcross})
and (\ref{rightcocross}) we obtain the cross product and coproduct structures
as stated. The isomorphism (\ref{cotheta}) with the dual of the double comes
out as
\eqn{matcotheta}{{\hat\theta}(L^i{}_j\tens 1)=\sum e_a\la L^i{}_j\tens
f^a,\quad {\hat\theta}(1\tens t^i{}_j)=Sl^+{}^i{}_k\tens t^k{}_j.}

Note finally that $R$ is no longer of real type: For the $sl_2$ case at
$|q|=1$, it obeys
\eqn{antirealRmat}{ \overline{R^i{}_j{}^k{}_l}=R^{-1}{}^i{}_j{}^k{}_l.}
Nevertheless, from Example~2.3 and the general theory above, we know that we
have a $*$-algebra cross product.
One can also verify this example explicitly, as follows. Firstly, the
$*$-structure on $U_q(sl(2,\R))$ takes the form
$l^\pm{}^i{}_j{}^*=q^{i-j}l^\pm{}^i{}_j$. From the pairing as a Hopf
$*$-algebra with $A(R)$ one obtains likewise $t^i{}_j{}^*=q^{i-j} S^2 t^i{}_j$
for the relevant $*$-structure for $SL_q(2,\R)$ (any even power of $S$ will do
here for a Hopf $*$-algebra, but this is the one that we need). This comes out
as
\[ \pmatrix{t^1{}_1{}^*&t^1{}_2{}^*\cr
t^2{}_1{}^*&t^2{}_2{}^*}=\pmatrix{t^1{}_1&qt^1{}_2\cr q^{-1}t^2{}_1&t^2{}_2}.\]
Next, the action $L_1\ra t_2=R^{-1} L_1 R$ was already computed in a slightly
more general context (for the degenerate Sklyanin algebra) in \cite{Ma:skl} as
$(\ )\ra t^1{}_2=0$ and
\[ \pmatrix{q^{H\over 2}\cr q^{-{H\over 2}}\cr X_+\cr X_-}\ra
t^1{}_1=\pmatrix{q^{H\over 2}\cr q^{-{H\over 2}}\cr q X_+\cr q^{-1}X_-},\
\pmatrix{\ \cr \cr }\ra t^1{}_1=\pmatrix{q^{H\over 2}\cr q^{-{H\over 2}}\cr
q^{-1} X_+\cr qX_-},\ \pmatrix{\ \cr \cr }\ra t^2{}_1=\pmatrix{\lambda(1-q)
X_+\cr \lambda (1-q^{-1})X_+q^{-H}\cr \lambda (1-q^{-1})X_+^2 q^{-{H\over
2}}\cr \lambda q^{-{H\over 2}}(X_+X_--qX_-X_+)}\]
where $\lambda=q^{-{1\over 2}}(q-q^{-1})$.
The non-trivial part of the verification of (\ref{rightuni}) is then
\[\left( \pmatrix{q^{H\over 2}\cr q^{-{H\over 2}}\cr X_+\cr X_-}\ra
t^2{}_1\right)^*=\pmatrix{\lambda(q-1) X_+\cr \lambda (q^{-1}-1)X_+q^{-H}\cr
\lambda (1-q^{-1})X_+^2 q^{-{H\over 2}}\cr \lambda q^{-{H\over
2}}(X_+X_--qX_-X_+)}=\pmatrix{q^{H\over 2}\cr q^{-{H\over 2}}\cr -X_+\cr
-X_-}\ra(-t^2{}_1)=
\pmatrix{q^{H\over 2}\cr q^{-{H\over 2}}\cr X_+\cr X_-}^*\ra(St^2{}_1)^*.\]
\endproof

This $*$-cross product is dual as a Hopf algebra to the quantum double model at
the end of Section~3. Nevertheless, we see that it has a similar interpretation
although, this time, $q$ is required to be of modulus 1 rather than real as
before. To complete this picture it remains only to fill out the details of
this interpretation.

Firstly, because $U_q(sl_2)$ is factorizable, we have an isomorphism
$BSL_q(2)\isom BU_q(sl_2)$. The latter is equipped now with a $*$-structure
inherited from $U_q(sl(2,\R))$ and this means that $BSL_q(2)$ also has a
$*$-structure. From the explicit form of $l^+Sl^-$\cite{FRT:lie} in this case,
\[\pmatrix{a&b\cr c&d}=\pmatrix{q^{H}& q^{-\h}(q-q^{-1})q^{H\over 2}X_-\cr
q^{-\h}(q-q^{-1})X_+q^{H\over 2}& q^{-H}+q^{-1}(q-q^{-1})^2X_+X_-}\]
one has
\eqn{*bsl2r}{ \pmatrix{a^*&b^*\cr c^*& d^*}=\pmatrix{a&q^2b\cr q^2 c & q^2
d+(1-q^2)a}.}
We denote the braided group $BSL_q(2)$ with this $*$-structure by $BSL_q(2,\R)$
in honour of the limit here as $q\to 1$. Its algebra relations are as in
(\ref{bm2a})-(\ref{bm2-det}). This gives our interpretation of the position
observables in Example~4.7.

Secondly, we want to view the momentum quantum group $SL_q(2,\R)$ as a quantum
enveloping algebra. This is not a new idea except that it is usually considered
with the compact $*$-structure or without consideration of the $*$-structure at
all. The algebra here is the usual $SL_q(2)$ one\cite{Dri}\cite{Wor:twi}. If we
define
\eqn{su*gen}{ q^\xi=t^1{}_1,\quad q^\eta=t^2{}_2,\quad \zeta={t^1{}_2\over
q-q^{-1}},\quad \chi={t^2{}_1\over q-q^{-1}}}
where we suppose that $q$ is generic, then the algebra relations become
\eqn{su*com}{ [\chi,\xi]=\chi=[\eta,\chi],\quad
[\zeta,\xi]=\zeta=[\eta,\zeta],\quad [\chi,\zeta]=0,\quad q^\xi
q^\eta=1+(q-q^{-1})^2q^{-1}\zeta\chi}
while its usual matrix coproduct becomes
\eqn{su*coproda}{ \Delta q^\xi=q^\xi\tens
q^\xi+(q-q^{-1})^2\zeta\tens\chi,\quad \Delta q^\eta=q^\eta\tens
q^\eta+(q-q^{-1})^2 \chi\tens\zeta}
\eqn{su*coprodb}{\Delta\zeta=\zeta\tens q^\eta+q^\xi\tens \zeta,\quad
\Delta\chi=\chi\tens q^\xi+q^\eta+\chi.}
Finally, the $*$-structure for $SL_q(2,\R)$ used in Example~4.7 becomes
\eqn{*su*}{ \xi^*=-\xi,\quad \eta^*=-\eta,\quad \chi^*=-q^{-1}\chi,\quad
\zeta^*=-q\zeta.}

We assume that $q=e^{t}$ (where $t$ in our case is imaginary) and deduce from
the Campell-Baker-Hausdorf formula applied to these equations that
\[   [\xi,\eta]=O(t),\quad \eta=-\xi+O(t).\]
The scaling of the generators is critical here and means that (with scaling as
defined) we have in the limit the Lie algebra
\eqn{su*lie}{ [\xi,\chi]=\chi,\quad [\xi,\zeta]=\zeta,\quad [\chi,\zeta]=0}
with its usual linear coproduct. This real Lie algebra is the solvable one
appearing in the Iwasawa decomposition of $sl_2$ and can be called the Drinfeld
dual $su_2^*$ of $su_2$ in view of Drinfeld's general theory of Lie
bialgebras\cite{Dri:ham}. Details of the computation from the Iwasawa
decomposition can be found for example in \cite[Sec. 2]{Ma:mat}. In our case we
see that the momentum quantum group in Example~4.7 can be regarded as a
q-deformation $U_q(su_2^*)$ of its enveloping algebra, cf the ideas introduced
in \cite[Sec. 7]{Dri}.

Finally, the right action of this q-momentum group on the co-ordinate
generators of $BSL_q(2,\R)$ is
\eqn{su*action}{ \pmatrix{a&b\cr c&d}\ra \xi=\pmatrix{0&-b\cr -c&0},\quad (\
)\ra\zeta=0,\quad \pmatrix{a&b\cr c&d}\ra \chi=\pmatrix{-qc&-(d-a)\cr 0 &
q^{-1}c}.}
In the limit $q\to 1$ we see that $\xi,\chi,\zeta$ become anti-self-adjoint and
have a $*$-representation on the algebra of functions on $SL(2,\R)$. This is
the system whose quantization as a $*$-algebra semidirect product is the dual
of the quantum double $D(U_q(sl(2,\R)))$ at least for generic $q$.

This completes the details of the dual model discussed in the Introduction in
(\ref{codouble.int}). Clearly, the physical meaning of these  quantum systems
is less familiar than those of the previous section. It is interesting however,
that these dual systems have a mathematical interpretation as a $*$-algebra
cross product provided $\CR$ is antireal, whereas in Section~3 it was required
to be real. It seems that both cross products are not quantum $*$-algebras
precisely at the same time, unless $\CR$ is triangular (in which case the
notions of real and antireal coincide).

\section{Concluding Remarks}

For completeness we conclude by placing the above results in the context of two
other interpretations of the quantum double (and of the semidirect products of
the type above) that are developed elsewhere. These are semidirect products as
a process of bosonization of braided objects\cite{Ma:bos} and semidirect
products as quantum principal bundles\cite{BrzMa:gau}.

The idea behind \cite{Ma:bos} is that a Hopf algebra in a braided category is
analogous to the idea of a super or $\Z_2$-graded structure, with the role of
the $\Z_2$ played by the quantum group $H$ that generates the braided category
(as explained in the Preliminaries). Because of the grading there is necesarily
a $\Z_2$-action and it is natural to `bosonize' the super-Hopf algebra $B$ into
an ordinary Hopf algebra $B\cocross \Z_2$. The idea is that the information
previously in the grading or bose-fermi statistics is used expressed as
non-commutativity in the algebra by adding an additional
generator $g$ with
\[ bg=(-1)^{\deg(b)}gb, \qquad g^2=1.\]
This trick is well-known to physicists under the heading of the Jordan-Wigner
transform\cite{MacMa:str} and also to mathematicians e.g.\cite{Fis:sch}. This
is also the idea behind \cite{Ma:bos} where we showed that every braided-Hopf
algebra in the category $\Rep(H)$ leads to an ordinary Hopf algebra $B\cocross
H$, its {\em bosonization}.

{}From this point of view, $D(H)$ is nothing other than the bosonization of the
braided-group $\und A$ of function algebra type associated to the
quasitriangular Hopf algebra $H$. Here $H$ is understood as generating the
braiding or braid-statistics under which the braided-group is covariant. Thus
two ideas, of grading and of momentum-covariance with associated Mackey
quantization are unified when both are viewed in the general context of quantum
groups. This gives insight into the nature of quantization as a process of
braiding, and is explored further in \cite{Ma:csta}.

One of the themes above has been the interaction of our constructions with Hopf
algebra duality. Here we want to mention a powerful diagrammatic way of making
such dualizations which is indispensable in a braided-group
context\cite{Ma:introp} but useful even for Hopf algebras. The point is that
the possibility of a dual Hopf algebra is based on the fact that the axioms of
a Hopf algebra have an input-output symmetry in which the axiom system, when
written as commuting diagrams, is invariant under reversal of the arrows. In
the diagrammatic notation one goes further and writes all maps as nodes on
strings flowing from the inputs down to the outputs. For example in
\cite{Ma:bos} we gave the proofs of the bosonization construction in this way.
Hence for the dual theorem one simply turns the diagram-proofs up-side-down.
This turns left modules into right comodules etc. and recovers the general
constructions of Section~4 directly. From this point of view the content of
Lemma~4.4 is precisely recovered as a dual version of the bosonization theorem:
if $C$ is a Hopf algebra living in the braided category of right comodules of a
dual-quasitriangular Hopf algebra then it has a dual-bosonization $C\cocross
A$. This diagrammatic view of Hopf algebra duality is important also in the
quantization interpretation where it suggests a kind of time-reversal and
parity invariance of the system.

Finally, its is known from a general theorem of Radford\cite{Rad:str} that
simultaneous products and coproducts of the type above are in one-to-one
correspondence with Hopf algebra projections. This means a Hopf algebra
surjection $\pi:P\to A$ say where $P,A$ are Hopf algebra and where the map
$\pi$ is split by a Hopf algebra inclusion $P{\buildrel i\over\hookleftarrow}
A$ in such a way that $\pi\circ i=\id$. If these quantum groups are like
functions on groups $G,H$ respectively then $\pi$ corresponds to an inclusion
$H\subset G$. In this case one can view $G\to G/H$ as a principal $H$-bundle.
In the same way, one can view $P\hookleftarrow B$ as a quantum principal bundle
with structure quantum group $A$ and base quantum space
\[  B=P^A=\{b\in P|\ (\id\tens\pi)\Delta b=b\tens 1\}.\]
In the case where $\pi$ is split the bundle is trivial with trivialization
provided by $i$. This means precisely that $P$ factorises as $P=B\cocross A$.
We refer to \cite{BrzMa:gau} for further details.

Thus we see that semidirect products such as we have studied above can equally
well be viewed as trivial quantum principal bundles. From this point of view,
the quantum double and its dual define two quantum bundles, the first with
structure quantum group $A=U_q(su_2)$ which we need to view as a quantum
function algebra of $SU(2)^*$ and the second with structure quantum group
$A=SL_q(2,\R)$. This time the perverse interpretation of an enveloping algebra
as function algebra occurs with the the model in Section~3 rather than the dual
model in Section~4.

These mutually dual interpretations of the quantum double as bundles parallel
then the mutually dual interpretations as quantization above. In summary, the
quantum double, as well as the more general cross products and coproducts,
allow us to extend the thesis of
\cite{Ma:phy}\cite{Ma:mat}\cite{Ma:hop}\cite{Ma:pla} that when both are
sufficiently generalized, quantization and geometry are the same thing,  from
mutually dual points of view.

\end{document}